\documentclass[twocolumn]{aastex61}
\usepackage{mathrsfs, mathtools}
\usepackage{captcont,xcolor}
\usepackage{blindtext}

\usepackage{CJKutf8}
\shortauthors{Long et al.}

\begin{document}
\begin{CJK*}{UTF8}{gbsn}

\title{Gaps and Rings in an ALMA Survey of Disks in the Taurus Star-forming Region}

\correspondingauthor{Feng Long}
\author{Feng Long(龙凤)}
\affiliation{Kavli Institute for Astronomy and Astrophysics, Peking University, Beijing 100871, China}
\affiliation{Department of Astronomy, School of Physics, Peking University,  Beijing 100871, China}

\author{Paola Pinilla}
\affiliation{Department of Astronomy/Steward Observatory, The University of Arizona, 933 North Cherry Avenue, Tucson, AZ 85721, USA}

\author{Gregory J. Herczeg(沈雷歌)}
\affiliation{Kavli Institute for Astronomy and Astrophysics, Peking University, Beijing 100871, China}
\author{Daniel Harsono}
\affiliation{Leiden Observatory, Leiden University, P.O. box 9513, 2300 RA Leiden, The Netherlands}

\author{Giovanni Dipierro}
\affiliation{Department of Physics and Astronomy, University of Leicester, Leicester LE1 7RH, UK}

\author{Ilaria Pascucci}
\affiliation{Lunar and Planetary Laboratory, University of Arizona, Tucson, AZ 85721, USA}
\affiliation{Earths in Other Solar Systems Team, NASA Nexus for Exoplanet System Science, USA}

\author{Nathan Hendler}
\affiliation{Lunar and Planetary Laboratory, University of Arizona, Tucson, AZ 85721, USA}

\author{Marco Tazzari}
\affiliation{Institute of Astronomy, University of Cambridge, Madingley Road, Cambridge CB3 0HA, UK}

\author{Enrico Ragusa}
\affiliation{Dipartimento di Fisica, Universita Degli Studi di Milano, Via Celoria, 16, I-20133 Milano, Italy}

\author{Colette Salyk}
\affiliation{Vassar College Physics and Astronomy Department, 124 Raymond Avenue, Poughkeepsie, NY 12604, USA}

\author{Suzan Edwards}
\affiliation{Five College Astronomy Department, Smith College, Northampton, MA 01063, USA}

\author{Giuseppe Lodato}
\affiliation{Dipartimento di Fisica, Universita Degli Studi di Milano, Via Celoria, 16, I-20133 Milano, Italy}

\author{Gerrit van de Plas}
\affiliation{Univ. Grenoble Alpes, CNRS, IPAG, F-38000 Grenoble, France}

\author{Doug Johnstone}
\affiliation{NRC Herzberg Astronomy and Astrophysics, 5071 West Saanich Road, Victoria, BC, V9E 2E7, Canada}
\affiliation{Department of Physics and Astronomy, University of Victoria, Victoria, BC, V8P 5C2, Canada}

\author{Yao Liu}
\affiliation{Max Planck Institute for Astronomy, K\"{o}nigstuhl 17, 69117 Heidelberg, Germany}
\affiliation{Purple Mountain Observatory, Chinese Academy of Sciences, 2 West Beijing Road, Nanjing 210008, China}

\author{Yann Boehler}
\affiliation{Univ. Grenoble Alpes, CNRS, IPAG, F-38000 Grenoble, France}
\affiliation{Rice University, Department of Physics and Astronomy, Main Street, 
77005 Houston, USA}

\author{Sylvie Cabrit}
\affiliation{Sorbonne Universit\'{e}, Observatoire de Paris, Universit\'{e} PSL, CNRS, LERMA, F-75014 Paris, France}
\affiliation{Univ. Grenoble Alpes, CNRS, IPAG, F-38000 Grenoble, France}

\author{Carlo F. Manara}
\affiliation{European Southern Observatory, Karl-Schwarzschild-Str. 2, D-85748 Garching bei M\"{u}nchen, Germany}

\author{Francois Menard}
\affiliation{Univ. Grenoble Alpes, CNRS, IPAG, F-38000 Grenoble, France}

\author{Gijs D. Mulders}
\affiliation{Department of the Geophysical Sciences, The University of Chicago, 5734 South Ellis Avenue, Chicago, IL 60637, USA}
\affiliation{Earths in Other Solar Systems Team, NASA Nexus for Exoplanet System Science, USA}

\author{Brunella Nisini}
\affiliation{INAF–Osservatorio Astronomico di Roma, via di Frascati 33, 00040 Monte Porzio Catone, Italy}

\author{William J. Fischer}
\affiliation{Space Telescope Science Institute Baltimore, MD 21218, USA}

\author{Elisabetta Rigliaco}
\affiliation{INAF-Osservatorio Astronomico di Padova, Vicolo dell'Osservatorio 5, 35122 Padova, Italy}

\author{Andrea Banzatti}
\affiliation{Lunar and Planetary Laboratory, University of Arizona, Tucson, AZ 85721, USA}

\author{Henning Avenhaus}
\affiliation{Max Planck Institute for Astronomy, K\"{o}nigstuhl 17, 69117 Heidelberg, Germany}

\author{Michael Gully-Santiago}
\affiliation{NASA Ames Research Center and Bay Area Environmental Research Institute, Moffett Field, CA 94035, USA}

\email{longfeng@pku.edu.cn}

\begin{abstract}
Rings are the most frequently revealed substructure in ALMA dust observations of protoplanetary disks, but their origin is still hotly debated. In this paper, we identify dust substructures in 12 disks and measure their properties to investigate how they form.  This subsample of disks is selected from a high-resolution ($\sim0.12''$) ALMA 1.33 mm survey of 32 disks in the Taurus star-forming region, which was designed to cover a wide range of sub-mm brightness and to be unbiased to previously known substructures. While axisymmetric rings and gaps are common within our sample, spiral patterns and high contrast azimuthal asymmetries are not detected. Fits of disk models to the visibilities lead to estimates of the location and shape of gaps and rings, the flux in each disk component, and the size of the disk.  The dust substructures occur across a wide range of stellar mass and disk brightness. Disks with multiple rings tend to be more massive and more extended.  The correlation between gap locations and widths, the intensity contrast between rings and gaps, and the separations of rings and gaps could all be explained if most gaps are opened by low-mass planets (super-Earths and Neptunes) in the condition of low disk turbulence ($\alpha=10^{-4}$). The gap locations are not well correlated with the expected locations of CO and N$_2$ ice lines, so condensation fronts are unlikely to be a universal mechanism to create gaps and rings, though they may play a role in some cases.  
\end{abstract}


\section{Introduction}     \label{sect:introduction}
Characterizing the structure of protoplanetary disks is crucial to understand the physical mechanisms responsible for disk evolution and planet formation. Given the typical size ($\sim$100 au) of protoplanetary disks (see review by \citealt{williams2011}), spatially resolving disks in nearby star-forming regions ($\lesssim$200 pc) requires observations with sub-arcsec resolution. Disk observations with the Atacama Large Millimeter/submillimeter Array (ALMA) have revealed a variety of disk substructures from thermal emission of mm-sized grains, dramatically changing our view of protoplanetary disks. Axisymmetric gaps and rings are the most frequently seen substructures, and have been observed in disks around HL Tau, TW Hydra, AA Tau, DM Tau, AS 209, Elias 2-24, V1094 Sco, HD 169142, HD 163296 and HD 97048 \citep{HLTau2015ApJ,andrews2016,isella2016,walsh2016,zhang2016,cieza2017,loomis2017,vanderplas2017,dipierro2018,fedele2018,vanTerwisga2018}, in both young and evolved systems around T Tauri and Herbig stars. Large azimuthal asymmetries also emerge in some systems \citep{brown2009,vanderMarel2013}, as well as spiral arms \citep{perez2016,tobin2016}. The origin of these substructures and their role in planet formation process are still widely debated.

In typical protoplanetary disks, mm-sized particles are expected to undergo fast radial drift towards the central star due to aerodynamic drag with the gas, resulting in severe depletion of mm-sized dust grains at large radii \citep{weidenschilling1977,birnstiel2014}. However, this picture is contradicted by high resolution images of mm-sized particles that are distributed over distances of tens or hundreds of au from the central star (see reviews by \citealt{testi2014} and \citealt{andrews2015}). Assuming that the rings revealed from ALMA are related to variations of dust density, the presence of rings indicates that inward drift of large dust grains (mm-sized) can be stopped or mitigated at specific radii. The physics that generates the rings therefore contributes to the persistence of mm-sized dust grains at large radii, even after a few Myr of disk evolution \citep[e.g.,][]{gonzalez2017}. The accumulation of dust in these regions might trigger efficient grain growth, thereby acting as an ideal cradle for forming planets \citep{carrasco-gonzalez2016}. A fundamental question then is what triggers the dust accumulation into ring shapes in disks and its connection to planet formation.

The mechanisms proposed to produce ring-like substructures in disks may be categorized into those related to disk physics and chemistry, and those related to planet-disk interactions.  When caused by disk physics and chemistry, the presence of a gap may trace the beginning of subsequent planet formation.  Some of the disk-specific mechanisms that can generate gaps and rings include: zonal flows induced by magneto-rotational instabilities \citep{johansen2009}, dead zones where gas accretion is regulated by spatial variations of the ionization level \citep{flock2015}, grain growth around condensation fronts \citep{zhang2015}, ambipolar diffusion-assisted reconnection in magnetically coupled disk-wind systems in the presence of a poloidal magnetic field \citep{suriano2018}, disk self-organization due to non-ideal MHD effects \citep{bethune2017}, suppressed grain growth with the effect of sintering \citep{okuzumi2016}, large scale instabilities due to dust settling \citep{loren-aguilar2016}, and secular gravitational instabilities regulated by disk viscosity \citep{takahashi2016}.

The disk gaps and rings could also be induced by interactions between the disk and planet(s) within the disk. On the one hand, a massive planet ($\gtrsim$ Neptune mass) embedded in the disk forms a gap in the gas density structure around its orbit, leading to the formation of a pressure bump outside the planet orbit, trapping large dust grains into rings and forming deep dust gaps (e.g., \citealt{lin1986,zhu2012,pinilla2012b}). On the other hand, a planet with a mass as low as 15 M$_\earth$ is able to slightly perturb the local radial gas velocity, inducing a ``traffic jam" that forms narrower and less depleted gaps \citep{rosotti2016}. Lower-mass planets can produce deep dust gaps without affecting the local gas structure \citep{fouchet2010,dipierro16a,dipierro17a}. Depending on the local disk conditions (e.g., temperature and viscosity) and planet properties, a single planet can also create multiple gaps \citep{bae2017,dong2017}.

Connecting these rings to the known distribution of exoplanets is challenging. Statistical studies of exo-planets reveal a higher occurrence rate of giant planets around solar-type stars than M-dwarfs, while this trend is not seen for smaller planets (see a recent review by \citealt{mulders2018}). For more massive stars, the formation of the cores of giant planets is expected to be more efficient \citep{kennedy2008}.  The surrounding disks would then have more material to build more massive planets, as suggested by the stellar-disk mass scaling relation from recent disk surveys (e.g., \citealt{andrews2013,pascucci2016}). If gaps are carved by giant planets, then deeper and wider gaps should be more prevalent around solar-mass stars than around stars of lower mass, although this picture would be complicated by any mass dependence in disk properties (e.g., low mass planets can more easily open gaps in inviscid disks, \citealt{dong2017}). In the case of ice lines, gaps should form at certain locations determined by the disk temperature profile, which broadly scales with stellar luminosity.

The analysis of gap and ring properties with stellar/disk properties should help us to discriminate between these different mechanisms. However, the small number of systems observed at high-spatial resolution ($\sim0.1''$) to date limits our knowledge about the origins of disk substructures.  Moreover, the set of disks imaged at high resolution is biased to brighter disks, many with near/mid-IR signatures of dust evolution, and collected from different star-forming regions and thus environments.  These biases frustrate attempts to determine the frequency of different types of substructures, how these substructures depend on properties of the star and disk, and any evolution of substructures with time.

In this paper, we investigate properties of substructures in 12 disks, selected from a sample of 32 disks in the Taurus star-forming region that were recently observed at high resolution with ALMA. The paper is organized as follows. In \S~\ref{sect:observations}, we describe our ALMA Cycle 4 observations and sample selection for the 12 disks. In \S~\ref{sect:model}, we present modeling approach for disk substructures in the visibility plane and the corresponding model results. We then discuss in detail the stellar and disk properties for the 12 disks, and the possible origins for dust substructures from analysis of the gap and ring properties in \S~\ref{sect:discussion}. Finally, the conclusions of this work are summarized in \S~\ref{sect:conclusions}.

\begin{deluxetable*}{lccccclcc}[!t]
\tabletypesize{\scriptsize}
\tablecaption{ALMA Cycle 4 Observations\label{tab:ALMAObservations}}
\tablewidth{0pt}
\tablehead{
\colhead{UTC Date} & \colhead{Number} & \colhead{Baseline Range} & \colhead{pwv} & \multicolumn{3}{c}{Calibrators} & \colhead{On-Source} & \colhead{Targets}\\ 
\colhead{ } & \colhead{Antennas} & \colhead{(m)} & \colhead{(mm)} & \colhead{Flux} &  \colhead{Bandpass}  &  \colhead{Phase}  &  \colhead{Time (min)} &  \colhead{} 
}
\startdata
2017 Aug 27 & 47 & 21-3638 & 0.5 & J0510+1800 & J0510+1800 & J0512+2927 & 4 & MWC 480 \\
 &  &  &  &  &  & J0435+2532\tablenotemark{*} & 4 & CI Tau, DL Tau, DN Tau, RY Tau \\
 &  &  &  &  &  & J0440+2728 & 4 & GO Tau \\
 &  &  &  &  &  & J0426+2327 & 1.5 & IQ Tau \\
\hline
2017 Aug 31 & 45 & 21-3697 & 1.3 & J1107-4449 & J1427-4206 & J1058-8003 & 9--10 & CIDA 9, DS Tau\\
\hline
2017 Aug 31 -- Sep 2 & 45 & 21-3697 & 1.5 & J0510+1800  & J0423-0120 & J0426+2327 & 8.5 & FT Tau, UZ Tau E \\
 &  &  &  &  &  & J0435+2532 & 10 & IP Tau \\
\enddata
\tablecomments{The 12 disks discussed in this paper come from three observing groups, thus the observation setup for the remaining one group is not shown here. }
\tablenotetext{*}{The scheduled phase calibrator (J0426+2327) for these disks was observed at different spectral windows from the science targets, thus phase calibration cannot be applied from the phase calibrator to our targets. We used the weaker check source (J0435+2532) instead to transfer phase solutions.}
\end{deluxetable*}

\section{Observations and sample selection}     \label{sect:observations}

\subsection{Observations and Data Reduction}
Our ALMA Cycle 4 program (ID: 2016.1.01164.S; PI: Herczeg) observed 32 disks in the Taurus star-forming region in Band 6 (1.33 mm) with high-spatial resolution ($\sim0.12''$, corresponding to $\sim16$ au for the typical distance to Taurus). Targets were selected for disks around stars with spectral type earlier than M3, excluding duplication at high resolution in archival data, close binaries ($0.1''-0.5''$), and stars with high extinction ($A_V>3$ mag). Further details of the sample will be described in a forthcoming paper.

The 32 disks were split into four different observing groups based on their sky coordinates. All observations were obtained from late August to early September 2017 using 45-47 12-m antennas on baselines of 21$\sim$3697 m (15$\sim$2780 k$\lambda$), with slight differences in each group (see Table \ref{tab:ALMAObservations}). The ALMA correlators were configured identically into four separate basebands for each observation. Two basebands were setup for continuum observations, centered at 218 and 233 GHz with bandwidths of 1.875 GHz. The average observing frequency is 225.5 GHz (wavelength of 1.33 mm). The other two windows cover the two CO isotopologue lines
and will not be discussed in this paper.
On-source integration times were $\sim$4 min per target for one group with relatively bright disks and $\sim$10 min per target for the other three groups. Table \ref{tab:ALMAObservations} summarizes the details of observation setups in each group.

The ALMA data were calibrated using the Common Astronomy Software Applications (\textsc{CASA}) package \citep{McMullin2007}, version 5.1.1. Following the data reduction scripts provided by ALMA, the atmospheric phase noise was first reduced using water vapor radiometer measurements. The standard bandpass, flux, and gain calibrations were then applied (see Table~\ref{tab:ALMAObservations}). Based on the phase and amplitude variations on calibrators, we estimate an absolute flux calibration uncertainty of $\sim$10\%. Continuum images were then created from the calibrated visibilities with \textsc{CASA} task \textsc{tclean}. For targets with initial signal-to-noise ratio (S/N) $\ga$100 in the image, we applied three rounds of phase (down to the integration time) and one round of amplitude self-calibration. For targets with initial S/N\textless100, we applied only one round of phase and one round of amplitude self-calibration.  For two disks with  S/N\textless30, self-calibration was not applied. After each round of self-calibration, we checked the image S/N, and would cease the procedure when no significant improvement was measured in the S/N. A few disks had only two rounds of phase and one round of amplitude self-calibration.  Self-calibration led to 20--30\% improvements in S/N for most disks, and a factor of 2 improvement in S/N for the brightest disks.  
The data visibilities were extracted from the self-calibrated measurement sets for further modeling. The final continuum images were produced with Briggs weighting and a robust parameter of +0.5 in \textsc{tclean}, resulting in a typical beam size of $0.14''\times0.11''$, and a median continuum rms of 0.05 mJy beam$^{-1}$. These observations are not sensitive to emission larger than $\sim 1.3''$ (corresponding to $\sim 180$ au for the typical distance of Taurus region), which is set by the maximum recoverable scale of the chosen antenna configuration.

\subsection{Sample Selection}
\begin{figure*}[!t]
\centering
    \includegraphics[width=0.99\textwidth]{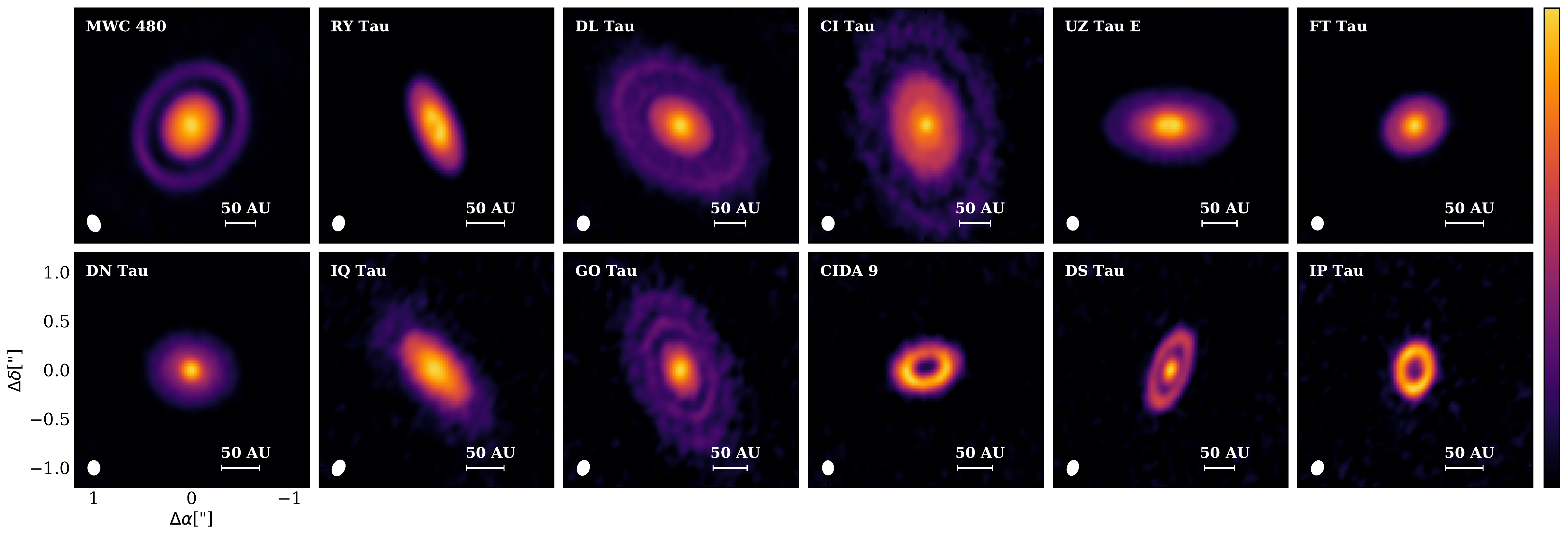} \\
    \caption{Synthesized images of the 1.33 mm continuum with a Briggs weighting of robust = 0.5. The images are displayed in order of decreasing mm flux, from the top left panel to the bottom right panel, and are scaled to highlight the weaker outer emission. The beam for each disk is shown in the left corner of each panel. \label{fig:cont_images}}
\end{figure*}

\begin{figure*}[!t]
\centering
    \includegraphics[width=0.99\textwidth]{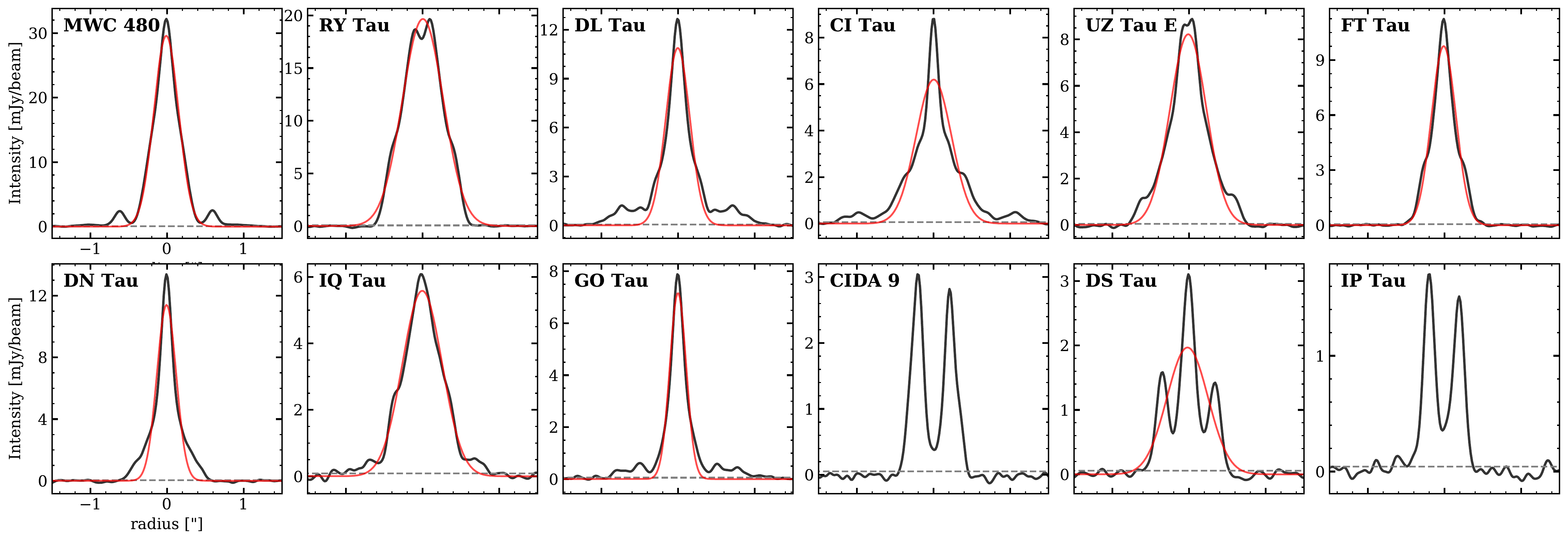} \\
    \caption{Radial intensity profiles (black lines) along disk major axis for the 12 selected disks with dust substructures, as the same order of Figure \ref{fig:cont_images}. The fitted Gaussian profile is shown in red to highlight the disk substructures, except for CIDA 9 and IP Tau, which have deep inner cavities. The $1\sigma$ noise level is shown in dashed line.  \label{fig:cont_rp_32}}
\end{figure*}

In this paper, we analyze the sub-sample of disks within our program that show prominent substructures in their dust thermal emission (see dust continuum images and radial profiles for the sub-sample in Figure~\ref{fig:cont_images} and Figure~\ref{fig:cont_rp_32}). Results for the full sample will be presented in a forthcoming paper.

Our sample selection of disk substructures is mainly guided by inspection of the disk radial intensity profiles.  We first determine the disk major axis by using \textsc{CASA} task \textsc{imfit} to fit an elliptical Gaussian profile to the continuum emission in the image plane. The radial intensity profile along the major axis is then used for an initial classification of disk substructures, including
1) inner cavities; 2) extended emission at large radii; and 3) resolved rings or emission bumps. Twelve of our sample of 32 disks show substructures, with dust emission that cannot be fit with a single smooth central component.

This selection of disks with substructures is confirmed by quantifying the reduced $\chi^2$ of fits
of Gaussian profiles to the radial intensity profile along the disk major axis, within the central $1.5''$ of the centroid (refer to red lines in Figure \ref{fig:cont_rp_32}).  The disks selected for this paper have the largest $\chi^2$ values.  The choice to focus on twelve sources is somewhat arbitrary, but disks with even slightly lower $\chi^2$ values would include those with subtle deviations from a Gaussian profile that could be well fit with a single tapered power-law (see Long et al.~in prep). The source properties for the 12 selected disks are summarized in Table \ref{tab:source_prop}. 

\begin{deluxetable*}{lrrrrrccc}[!t]
\tabletypesize{\scriptsize}
\tablecaption{Source Properties and observation results\label{tab:source_prop}}
\tablewidth{0pt}
\tablehead{
\colhead{Name} & \colhead{SpTy} & \colhead{$T_{\rm eff}$} & \colhead{$M_*$} & \colhead{log($L_*$)} & \colhead{refs}  & \colhead{Distance} & \colhead{rms} & \colhead{Beam Size} \\
\colhead{} & \colhead{} & \colhead{(K)} &\colhead{($M_\odot$)} & \colhead{($L_\odot$)} &  \colhead{} & \colhead{(pc)} & \colhead{(mJy beam$^{-1}$)} & \colhead{(arcsec$\times$arcsec)} 
} 
\startdata
 CIDA 9   & M1.8 & 3589 & 0.43 & -0.7  &   HH14 & 171 &  0.05 & 0.13$\times$0.10\\
   IP Tau & M0.6 & 3763 & 0.52 & -0.47 &   HH14 & 130 & 0.047 & 0.14$\times$0.11\\
   RY Tau &   F7 & 6220 & 2.04 &  1.09 &   HH14 & 128 & 0.051 & 0.15$\times$0.11\\
UZ Tau E  & M1.9 & 3574 & 0.39 & -0.46 &   HH14 & 131 & 0.049 & 0.13$\times$0.11\\
   DS Tau & M0.4 & 3792 & 0.58 & -0.61 &   HH14 & 159 &  0.05 & 0.14$\times$0.10\\
   FT Tau & M2.8 & 3444 & 0.34 & -0.83 &   HH14 & 127 & 0.047 & 0.13$\times$0.11\\
  MWC 480 & A4.5 & 8460 & 1.91 &  1.24 & YLiu18 & 161 &  0.07 & 0.17$\times$0.11\\
   DN Tau & M0.3 & 3806 & 0.52 & -0.16 &   HH14 & 128 &  0.05 & 0.14$\times$0.11\\
   GO Tau & M2.3 & 3516 & 0.36 & -0.67 &   HH14 & 144 & 0.049 & 0.14$\times$0.11\\
   IQ Tau & M1.1 & 3690 & 0.50 & -0.67 &   HH14 & 131 & 0.076 & 0.16$\times$0.11\\
   DL Tau & K5.5 & 4277 & 0.98 & -0.19 &   HH14 & 159 & 0.048 & 0.14$\times$0.11\\
   CI Tau & K5.5 & 4277 & 0.89 & -0.09 &   HH14 & 158 &  0.05 & 0.13$\times$0.11\\
   \enddata
\tablecomments{The distance for each target is adopted from the Gaia DR2 parallax \citep{gaia2016,gaia2018}. Spectral type and stellar luminosity are adopted from the listed references (\citealt{herczeg2014} and Liu et al. submitted) and are updated to the new Gaia distance. Stellar masses are re-calculated with the stellar luminosity and effective temperature listed here using the same method in \citet{pascucci2016}. Further details will be described in a forthcoming paper of the full sample. }
\end{deluxetable*}

\section{Modeling disk substructures}\label{sect:model}
The 1.33 mm continuum images for our 12 disks (in Figure~\ref{fig:cont_images}) reveal substructures with a wide variety of properties.
Resolved rings are the most common type of substructures, characterizing half of our sample.  Several disks have two or more rings.  Emission bumps are detected from several disks, and would likely be resolved into clear rings with higher spatial resolution.  Four disks have inner disk cavities (surrounded by one or multiple rings), with different degrees of dust depletion and subtle azimuthal asymmetries. 
Spirals and high-contrast (with an intensity ratio higher than 2) azimuthal asymmetries are not seen in our sample.  These general results are consistent with expectations based on previous results of biased samples, which showed that rings are common while large azimuthal asymmetries (azimuthal dust traps, such as vortices) and spirals are rare.

In this section, we describe the general procedure in modeling the dust substructures performed in the visibility plane and present the results of best-fit models. Disk mm fluxes and disk dust sizes are then measured from the best-fit intensity profiles, which will be used in later analysis.

\subsection{Modeling Procedure} \label{sect:Model-Procedure}
In order to precisely quantify the observed morphology of dust continuum emission, our analysis is performed in the Fourier plane by comparing the observed visibilities to synthetic visibilities computed from a model intensity profile. Axisymmetry is assumed, since high-contrast asymmetries in the dust emission are not seen (Figure~\ref{fig:cont_images}; low-contrast asymmetries will be discussed briefly in \S~\ref{sect:asymmtry}).  Each disk is initially approximated by combining a central Gaussian profile with additional radial Gaussian rings, with the model intensity profile expressed as:
\begin{equation}
I(r) = A \exp \left( -\frac{ r^{2} }{2  \sigma_{0} ^{2} } \right) + \sum_i B_{i} \exp \left[ -\frac{ (r- R_{i} )^{2} }{2  \sigma_{i} ^{2} } \right]
\label{Eq:gaussian}
\end{equation}
where the first term represents the central emission and the second term represents a series of peaks in the radial intensity profile, and $R_{i}$ and $\sigma_{i}$ are the locations and widths of the emission components. In some cases, the central Gaussian profile is replaced with an exponentially tapered power-law, which reproduces the oscillation pattern in the visibility profile \citep{andrews2012,hogerheijde2016, zhang2016} and better fits the data (with two more free parameters). The revised model is then described as: 
\begin{equation}
I(r) = A {\left(\frac{ r }{ r_t}\right)}^{-\gamma} \exp\left[-\left(\frac{ r }{  r_t}\right)^{\beta}\right] + \sum_i B_{i} \exp\left[-\frac{ (r- R_{i} )^{2} }{2  \sigma_{i} ^{2} }\right]
\label{Eq:power}
\end{equation}
where $r_t$ is the transition radius, $\gamma$ is the surface brightness gradient index, and $\beta$ is the exponentially tapered index. The model visibilities are then created by Fourier transforming the disk model intensity profile using the publicly available code \textit{Galario} \citep{tazzari2018}. Fitting the model visibilities to the data visibilities is later performed with the \textit{emcee}\footnote{https://pypi.org/project/emcee/} package \citep{Foreman-Mackey2013}, in which a Markov chain Monte Carlo (MCMC) method is used to explore the optimal value of free parameters.

Our choice of component type and number in the model intensity profile for each disk is guided by the observed radial profile along the disk major axis (Figure \ref{fig:cont_rp_32}). 
A resolved ring or emission bump is modeled as a Gaussian ring component. 
The initial guesses for the amplitude, location and width of each component are also inferred from the radial profiles.  The disk inclination angle ($i$), the disk position angle (PA), and the position offsets from the phase center ($\Delta \alpha$ and $\Delta \delta$) are all free parameters in our fit. The starting point for the four parameters are estimated by fitting an elliptical Gaussian component to the continuum image with \textsc{CASA} task \textsc{imfit}. Prior ranges are set as $\pm$20\,deg for $i$ and PA, and $\pm$0.5\,arcsec for the position offsets. A uniform prior probability distribution is adopted for each of these parameters.

The radial grid in our model is linearly distributed within [0.0001$''$ - 4$''$] in steps of 0.001$''$, which is much smaller than our synthesized beam ($\sim 0.1''$).  We start the MCMC fit by exploring all free parameters (4 disk geometric parameters, plus Gaussian profile and Gaussian Ring(s)) with 100 walkers and 5000 steps for each walker. The burn-in phase for convergence is typically $\sim$2000 steps. A second run with parameter ranges confined from the initial run is conducted with another 5000 steps. For the second run, the autocorrelation time is typically 100 steps. The posterior distributions are then sampled using the chains of the last 1000 steps, as well as the optimal value (median value) and its associated uncertainty for each parameter. The statistical uncertainty for each parameter is estimated as the interval from the 16th to the 84th percentile.

In the next step, we perform multiple comparisons between data and model to check the goodness of our best-fit model, including visibility profiles, synthesized images, and radial cuts from the images. If significant symmetrical residuals ($\gtrsim5-10\sigma$) are present, we either include an additional Gaussian ring component or replace the central Gaussian profile with a tapered power-law.
These procedures are repeated until a reasonable best-fit is found.

\begin{figure*}[!t]
\centering
    \includegraphics[width=0.99\textwidth]{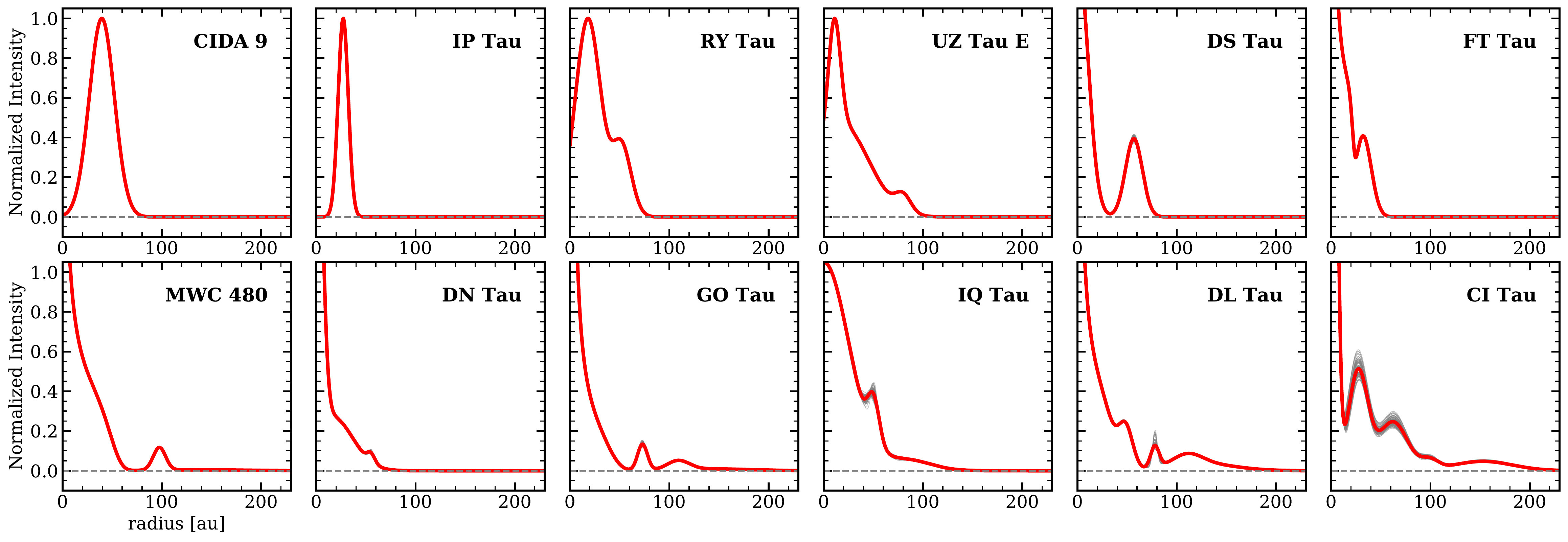}
    \caption{Best-fit intensity profiles (red line) from the MCMC fits, with 100 randomly selected models from the fitting chains overlaid in grey. For the four disks with inner cavities, the profiles are normalized to the peak of the ring.  For all other disks, the profiles are normalized to the values at 8 au to highlight the faint substructures in the outer disk. \label{fig:intensity_model}}
\end{figure*}

\begin{deluxetable*}{lrrrrccl}[!t]
\tabletypesize{\scriptsize}
\tablecaption{Disk Model Parameters\label{tab:fitting_model}}
\tablewidth{0pt}
\tablehead{
\colhead{Name} & \colhead{$F_\nu$} & \colhead{$R_{\rm eff}$} & \colhead{incl} & \colhead{PA} & \colhead{$\Delta \alpha$} & \colhead{$\Delta \delta$}  & \colhead{morphology/model description}\\
\colhead{} & \colhead{(mJy)} & \colhead{(au)} & \colhead{(deg)} & \colhead{(deg)} & \colhead{(arcsec)} &  \colhead{(arcsec)} & \colhead{}
} 
\startdata
CIDA 9&37.1$_{-0.09}^{+0.09}$&59.0$_{-0.17}^{+0.17}$&45.56$_{-0.18}^{+0.19}$&102.65$_{-0.26}^{+0.25}$&-0.51&-0.73 & inner cavity\\
IP Tau&14.53$_{-0.08}^{+0.07}$&34.58$_{-0.26}^{+0.13}$&45.24$_{-0.33}^{+0.32}$&173.0$_{-0.42}^{+0.43}$&0.05&0.17 & inner cavity\\
RY Tau&210.39$_{-0.1}^{+0.09}$&60.8$_{-0.13}^{+0.0}$&65.0$_{-0.02}^{+0.02}$&23.06$_{-0.02}^{+0.02}$&-0.05&-0.09 & inner cavity + 1 emission bump\\
UZ Tau E&129.52$_{-0.16}^{+0.14}$&81.61$_{-0.13}^{+0.13}$&56.15$_{-0.07}^{+0.07}$&90.39$_{-0.08}^{+0.08}$&0.77&-0.27 & inner cavity + 2 emission bumps\\
DS Tau&22.24$_{-0.11}^{+0.07}$&67.58$_{-0.32}^{+0.32}$&65.19$_{-0.13}^{+0.13}$&159.62$_{-0.14}^{+0.14}$&-0.13&0.22 & inner Gaussian profile + 1 ring\\
FT Tau&89.77$_{-0.1}^{+0.09}$&42.04$_{-0.13}^{+0.0}$&35.55$_{-0.16}^{+0.14}$&121.8$_{-0.27}^{+0.26}$&-0.1&0.13 & inner power-law profile + 1 emission bump\\
MWC 480&267.76$_{-0.21}^{+0.18}$&104.97$_{-0.16}^{+0.16}$&36.48$_{-0.05}^{+0.05}$&147.5$_{-0.08}^{+0.09}$&-0.01&0.0 & inner power-law profile + 1 ring\\
DN Tau&88.61$_{-0.2}^{+0.09}$&56.06$_{-0.13}^{+0.13}$&35.18$_{-0.22}^{+0.2}$&79.19$_{-0.38}^{+0.36}$&0.08&0.0 & inner Gaussian profile + 2 emission bumps\\
GO Tau&54.76$_{-0.2}^{+0.33}$&144.14$_{-2.45}^{+1.15}$&53.91$_{-0.2}^{+0.2}$&20.89$_{-0.24}^{+0.24}$&-0.17&-0.41 & inner power-law profile + 2 rings \\
IQ Tau&64.11$_{-0.34}^{+0.25}$&95.89$_{-1.05}^{+0.66}$&62.12$_{-0.2}^{+0.19}$&42.38$_{-0.23}^{+0.22}$&-0.09&0.07 & inner Gaussian profile + 2 emission bumps \\
DL Tau&170.72$_{-0.16}^{+0.37}$&147.39$_{-0.16}^{+0.48}$&44.95$_{-0.09}^{+0.09}$&52.14$_{-0.14}^{+0.15}$&0.24&-0.06 & inner power-law profile + 1 emission bump + 2 rings \\
CI Tau&142.4$_{-0.24}^{+0.15}$&173.8$_{-0.32}^{+0.47}$&49.99$_{-0.12}^{+0.11}$&11.22$_{-0.13}^{+0.13}$&0.33&-0.08 & inner Gaussian profile + 3 emission bumps + 1 ring\\
\enddata
\tablecomments{The inclination, PA, and phase center offsets ($\Delta \alpha$ and $\Delta \delta$) are parameters fitted with MCMC. Total flux ($F_\nu$) and effective radius ($R_{\rm eff}$, with 90\% flux encircled) are derived from the best-fit intensity profile for each disk. The quoted uncertainties are the interval from the 16th to the 84th percentile of the model chains. The typical uncertainties for $\Delta \alpha$ and $\Delta \delta$ are $<0.001''$, thus not listed. An emission bump or an resolved ring is modeled by a Gaussian ring. The faint outer ring for DL Tau and GO Tau is included to describe the tenuous outer disk, and the faint $3\sigma$ ring for MWC 480 is indicated from the fitting residual map. The three faint outer rings are not included in the description column and will not be used in the analysis in \S~\ref{sect:discussion}. }
\end{deluxetable*}

\subsection{Modeling Results} \label{sect:Model-Result}
Detailed results of the best-fit models are presented here, as well as the approach to derive total disk flux and disk size based on the best-fit models. Our final choice of the best-fit model for each disk is guided by using the fewest number of parameters to reproduce the axisymmetric structures with residuals less than $\sim 5\sigma$.  Figure~\ref{fig:model_result_all} in the Appendix compares the best-fit model with the observed visibility profiles, synthesized images, and radial profiles for each disk. In general, our models fit the disk total flux and disk substructures reasonably well, as indicated from the consistency of data and model visibilities at the shortest baseline and the match of visibility structures at the longer baselines, respectively. For disks with azimuthal asymmetries, our model fail to accurately reproduce the amplitude of the substructure component, but captures the location and width of the ring(s) well, which are the main focus of our analysis below.

\subsubsection{Best-Fit Models}
The best-fit model intensity profiles for the 12 disks are shown in Figure \ref{fig:intensity_model}, with the substructure component types and numbers of each disk summarized in Table \ref{tab:fitting_model}. The detailed information (e.g., gap and ring location/width) are provided in Appendix Table~\ref{tab:gap_info}.

The inner regions of four disks are described with a Gaussian profile (Eq.~\ref{Eq:gaussian}), four disks are described by a revised power-law model (Eq.~\ref{Eq:power}), and the four other disks lack mm-emission from their inner disks (see Table \ref{tab:fitting_model} for details).  
For the inner disks described by a power law, the taper index $\beta$ ($>$4) corresponds to a sharp outer edge of mm-sized dust for the emission of the inner blob, consistent with the prediction of fast radial drift of dust particles \citep{birnstiel2014}.

The four disks with inner cavities are fit with (a sum of) Gaussian ring(s). In three disks (MWC 480, GO Tau, and DL Tau), an additional Gaussian ring in the outermost disk is included to account for the tenuous outer disk edge, which is detected at $\sim 3\sigma$ significance. The inclusion of one more component for GO Tau and DL Tau is needed to avoid generating an outer ring with a width that is much broader than observed.\footnote{Instead of adding another Gaussian ring to describe the tenuous outer disk, we test with a Nuker profile, which could produce an asymmetric ring \citep{tripathi2017}. The derived gap and ring properties are consistent within uncertainties in two models.}  The additional ring for MWC 480 at 1$''$ radius is needed to reproduce the $3\sigma$ ring in the residual map, which is found in the fitting of the visibilities, but is too faint to be visible in the observed image. The modeling of MWC 480 disk by Liu et al.~(submitted) does not include this component, since they start the modeling in the image plane and focus on reproducing the primary structures. 
The detailed analysis of the substructure components will be presented in Section \ref{sect:discussion}.

The disk geometry parameters are summarized in Table~\ref{tab:fitting_model}, in which the best-fit inclination and position angles are generally consistent with the values estimated from \textsc{imfit} within 2-3$\degr$. The largest difference of PA is seen in DN Tau, in which our best-fit PA is 6$\degr$ larger (to the east) than the initial \textsc{imfit} estimation, hinting for some difference in disk orientation between the emission of the inner blob and the outer ring (see also the 3$\sigma$ residual in Figure \ref{fig:model_result_all}). The differences between the inclinations and position angles from simplistic models in the image plane with \textsc{imfit} and those from our visibility fitting suggest that the formal errors listed in Table 3 are likely underestimated.

\subsubsection{mm Flux and Dust Disk Size}
The disk flux densities at 1.33 mm and dust disk sizes are inferred from the model intensity profile, as described in this subsection, and are not model parameters that are directly fit in MCMC.
 Disk mm fluxes and disk dust sizes are summarized in Table \ref{tab:fitting_model}.

Given an intensity profile, the cumulative distribution could be described as,
\begin{equation}
f_\nu(r) = 2\pi \int_0^r I_\nu(r')r'dr',
\end{equation}
thus the total flux is $F_\nu = f_\nu(\infty)$ by definition. The mm flux for each disk is measured by integrating over the best-fit intensity profile. We then randomly choose 100 models in the last 1000 steps ($\times$ 100 walkers) of our MCMC chain to estimate flux uncertainty as the central interval from 16th to 84th percentile. For most disks, our flux measurements at 1.33 mm are consistent with pre-ALMA interferometry measurements\footnote{Flux densities at 1.33 mm in  \citet{andrews2013} are determined from power-law fits, where $F_\nu \propto \nu^\alpha$, by using all available measurements in the literature in the 0.7--3 mm wavelength range.}
within uncertainties \citep{andrews2013}, assuming 10\% and 15\% absolute flux uncertainty for ALMA and pre-ALMA results, respectively. Our flux densities for CI Tau, FT Tau, and IP Tau are more than $30\%$ brighter than those reported in \citet{andrews2013}.
However, the measured flux density for CI Tau is highly consistent with a recent ALMA measurement \citep{konishi2018}, and the FT Tau flux density is similar to a past CARMA measurement \citep{kwon2015}. For IP Tau, the flux difference is reconciled if the SMA measurement at 0.88 mm is extrapolated to 1.33 mm with a spectral index of 2.4 \citep{andrews2013,tripathi2017}.  These modest inconsistencies are likely related to unknown systematic flux calibration uncertainty, self-calibration, and different methods in estimating fluxes. These differences in fluxes will not affect the results in our following analysis.

The effective disk radius, $R_{\rm eff}$, is defined here as the radius where 90\% of the total flux is encircled \citep[see, e.g.][]{tripathi2017}.  The uncertainty for disk size is estimated in the same way as the flux uncertainty. We do not compare the disk sizes with results in \citet{tripathi2017} for the few overlapping disks, since the two works probe different wavelengths and use different size metrics.

\subsection{Residuals and Azimuthal Asymmetries} \label{sect:asymmtry}
The best fits to the observed visibilities yield significant residuals ($>10\sigma$) for a few disks. These residuals indicate azimuthal asymmetries for the innermost rings of CIDA 9, RY Tau, and UZ Tau E. One characteristic feature of this set of disks is that their inner regions are depleted of dust (including marginal depletions). High contrast asymmetries have been observed in some transition disks  (e.g., IRS 48 by \citealt{vanderMarel2013}), and interpreted as vortices that could be triggered by the presence of planets. An eccentric cavity, induced by companions in the inner disk, could also create azimuthal asymmetries, with contrast levels depending on the mass of the companion \citep{ataiee2013,ragusa2017}.
The azimuthal asymmetry of the AA Tau disk has alternatively been attributed to a misalignment between the inner and outer disks \citep{loomis2017}. 
Additional azimuthal structures on top of an underlying axisymmetric disk model would be needed to better describe the emission pattern, and is beyond the scope of this paper.

The inner emission blob of CI Tau and GO Tau also return modest residuals of $\sim5\sigma$. These inner emission regions have a narrow extent of 0.1$''$-0.2$''$ in radius, so subtle radial variations might be present but are not well enough resolved to interpret here. The hot-Jupiter candidate around CI Tau found by  \citet{johns-Krull2016} is in a 9-day orbit and likely does not affect the rings detected here on much larger radii.

\begin{figure*}[!t]
\centering
    \includegraphics[width=0.45\textwidth]{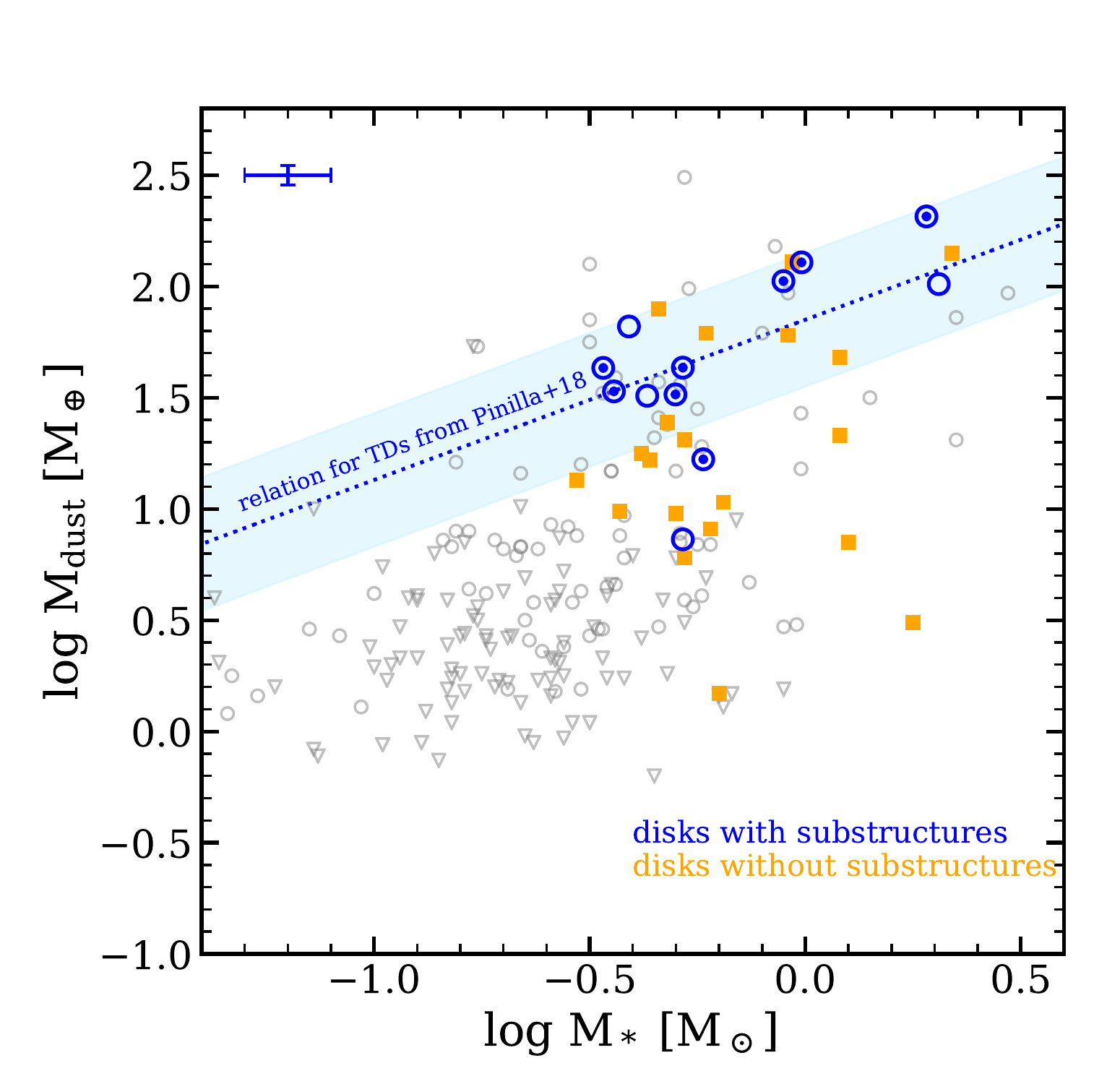}
    \includegraphics[width=0.45\textwidth]{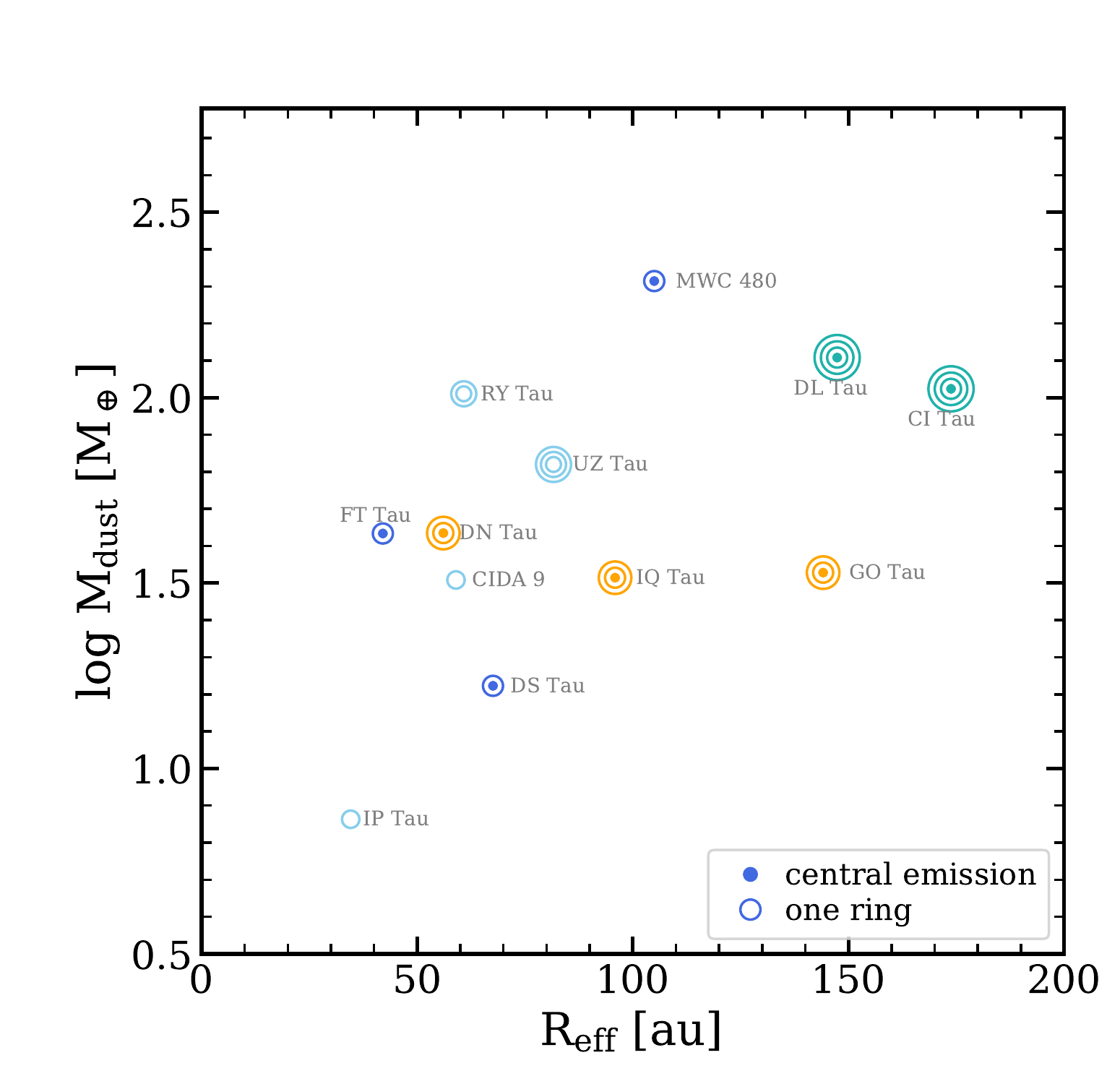}
    \caption{Left: stellar mass versus disk mass for the 12 disks with substructures (in blue, open circles for the four disks with inner cavities), the 20 disks without substructures in current observations (in orange, using disk masses from \citealt{andrews2013}), and the full Taurus sample (in grey, upper limits in triangles) of \citet{andrews2013}. The relationship between stellar mass and disk mass for transition disks (dotted blue line) is taken from \citet{pinilla2018}, with shaded region showing the typical data scatter. The typical error in $\log$($M_{\rm dust}$) of $\sim 0.04$ dex, including the 10\% flux calibration uncertainty, and in $\log$($M_*$) of 0.1 dex, are shown in the left corner; Right: disk effective radius versus disk dust mass for the 12 disks with substructures, with colors and symbol shapes separating disks with single, double, and multiple rings and  disks with inner cavities.  \label{fig:Md-Ms-R}}
\end{figure*}

\section{Results and Discussion}		\label{sect:discussion}
Previous measurements of disk substructures have been biased to brighter disks or disks in which the presence of substructures have already been inferred from other observations. The disk substructures identified in this survey are seen for the first time \footnote{In a contemporaneous paper, \citet{clarke2018} used higher-resolution ALMA images of the CI Tau disk to identify three prominent gaps, with properties that are broadly consistent with the three gaps measured in our coarser data.} at $\sim 0.1''$ resolution in an unbiased study that covers a wide range in fluxes within a given range of stellar mass.

From our full sample of 32 disks, we have identified 12 disks with substructures in their dust continuum emission.  Four disks have inner cavities in the mm continuum, encircled by single rings for CIDA 9 and IP Tau and multiple rings for RY Tau and UZ Tau E.  Three disks (FT Tau, DS Tau, and MWC 480), have mm continuum emission characterized by an inner disk encircled by a single ring.  Five disks (CI Tau, DL Tau, GO Tau, IQ Tau, and DN Tau), have an inner disk encircled by multiple rings.  The location and shape for each of these components are modeled as symmetric Gaussian profiles and are fit in the visibility plane (\S~\ref{sect:Model-Procedure}).

Table~\ref{tab:gap_info} in the Appendix summarizes the results from our fits, including the size of inner cavities, the radial location and width of gaps and rings, and the flux contrast ratio between the rings and gaps.  The properties of the substructures are disparate, with radial locations from 10--120 au, rings with emission that accounts for 10--100\% of the total flux from the disk, and widths that are usually $\sim$ 0.2 times the radial location of the gap, but can be wider.  The presence of most of these substructures does not obviously depend on any disk or stellar property.

In this section, we synthesize these disparate properties in an attempt to identify the physical mechanism(s) that produce cavities and rings.  We begin by exploring the parameter space occupied by our sample to describe the star and disk properties of our substructures. We then apply our results to expectations for the properties of gaps and rings that could be introduced by condensation fronts and by planets. The bulk of gaps could be carved by planets with masses close to the minimum planet mass able to produce gas pressure bumps, while less than half of the gaps are close to volatile condensation fronts.


\subsection{Source Properties for Disks with Substructures}
Substructures in our sample are present in objects that cover a wide range in stellar and disk mass. The left panel of Figure~\ref{fig:Md-Ms-R} shows the location of our 12 disks with dust substructures in the $M_* - M_{\rm dust}$ plane, as well as the other 20 disks in our full sample, which do not show dust substructures at our current resolution. The full Taurus sample from \citet{andrews2013} is also included in this plot to provide a broader comparison.  The dust masses are estimated from the 1.33 mm continuum flux density (e.g., \citealt{beckwith1990}) by
\begin{equation}
M_{\rm dust} = \frac{D^2 F_\nu}{\kappa_\nu B_\nu(T_{\rm dust})},
\end{equation}
with a dust opacity $\kappa_\nu=2.3\,\rm{cm}^2\,\rm{g}^{-1} \times (\nu/230\,\rm{GHz})^{0.4}$, a Planck function $B_\nu(T_{\rm dust})$ for a dust temperature of 20 K for each disk at distance $D$, assuming the dust is optically thin.
Stellar effective temperatures and stellar luminosities are adopted from \citet{herczeg2014} and then updated for individual Gaia DR2 distances \citep{gaia2018} for the full sample.  The stellar masses are then calculated from the \citet{baraffe15} and non-magnetic \citet{feiden16} evolutionary tracks, following \citet{pascucci2016}. The disk dust masses are calculated with updated Gaia DR2 distance for individual object with mm fluxes adopted from our measurements for the 12 disks and adopted from \citet{andrews2013} for the other Taurus members. 
The uncertainties of our estimated dust masses only consider the uncertainties of flux measurements, and do not take into account the differences in dust temperature and dust optical depth.

Our sample focuses on disks around  M-dwarfs and solar-mass stars in Taurus, with a requirement that the spectra type of the star is earlier than M3 (corresponding to $\sim 0.25$ M$_\odot$ in the \citet{baraffe15} evolutionary tracks and $\sim 0.45$ M$_\odot$ in the magnetic \citet{feiden16} tracks, for an age of 2 Myr). Disks with dust substructures cover this full stellar mass range of our whole sample.  The two disks around early spectral types (the A4 star MWC 480 and the F7 star RY Tau) both have prominent dust rings.
Disk dust masses for our 12 disks scatter over more than one order of magnitude, even in a narrow stellar mass bin. Dust substructures seem to be more common in brighter disks.  
A more complete analysis of the statistics with respect to the parent sample of 32 objects will be presented in a forthcoming paper.

Four disks in our sample have resolved inner cavities (including marginal depletion), including three new discoveries and confirmation of the inner cavity of RY Tau found by \citet{pinilla2018}.  None of these four disks show any signature of a cavity based on the SED (see also Figure 13 in \citealt{andrews2013}), so they all have warm dust near the star, similar to some of the cavities found by \citet{andrews11}.
In an analysis of 29 disks with inner mm cavities observed by ALMA, \citet{pinilla2018} found a flatter $M_* - M_{\rm dust}$ relation when compared to the correlation obtained for all disks from several different star-forming regions.  Inner cavities may therefore be more common among more massive disks, regardless of stellar mass. 
Three of our four inner cavity disks are consistent with this correlation, and are in the upper end of masses for all Taurus disks.  The exception, IP Tau, has the smallest and faintest disk in our sample.
Some outliers may be expected from this relationship for circumbinary disks, such as the disk around CoKu Tau 4 \citep{dalessio05,ireland08}. With a cavity radius of 21 au, a companion to IP Tau would need to be located at $\sim 10$ au, or $\sim0\farcs06$ \citep{artymowicz94}.  Previous binary searches would have been unable to resolve such a close companion.  Unlike CoKu Tau 4, IP Tau must have an inner disk to explain the IR excess \citep{furlan06} and active, though very weak accretion \citep{gullbring98}. 
Future high angular resolution and sensitivity observations of faint disks will test whether the \citet{pinilla2018} relationship is robust to the selection bias that past ALMA observations (Cycle 0 to Cycle 3) had towards brighter disks.

For the four disks with inner cavities, only one ring is detected from IP Tau and CIDA 9, while two rings are detected from RY Tau and three from UZ Tau E. The outer substructures of RY Tau and UZ Tau E are not clearly seen in the images but are detected in the uv-plane and in the cross-cut of the image along the semi-major access in Figure~\ref{fig:cont_rp_32}.  In contrast, the disks of IP Tau and CIDA 9 have large inner cavities, with faint dust emission detected at low S/N (see the right panel of Figure~\ref{fig:Md-Ms-R}) that suggests larger depletion of mm grains. However, the outer substructures of RY Tau and UZ Tau E would still have been detected if the S/N were scaled down to match the fainter signal of IP Tau and CIDA 9.  One possibility is that narrow cavities, as in the case of RY Tau and UZ Tau E\footnote{UZ Tau E is a spectroscopic binary with a separation of $\sim 0.03$ au \citep{mathieu1996}, which is far too tight to create the $\sim 10$ au cavity \citep{artymowicz94}.}, will evolve into larger and more depleted cavities. Nevertheless, more observations of inner cavities disks spanning different cavity sizes, ages, and dust depletion factors are required to test this hypothesis.

As shown in the right panel of Figure~\ref{fig:Md-Ms-R}, the number of substructures (single, double, or multiple rings) seems independent of whether mm continuum emission is present in the inner disk.  Disks with similar brightness have dust distributions that are diverse in shape and in size.
However, disks with multiple rings tend to be brighter and more extended, implying that substructures that originate at larger radii may act as mechanisms (e.g., dust traps) preventing the loss of mm-sized dust due to inward drift, thus retaining both the high dust mass and large disk size (e.g., \citealt{pinilla2012}). 
This explanation should be valid if $R_{\rm gas}/R_{\rm dust}$ is higher in smaller disks, for which dust inward migration is very efficient without ``traps" formed at larger radii, though this is not seen in \citet{ansdell2018}.  
Alternatively, the initial distribution of disk sizes in young stellar objects may be bimodal, with some large and some small, depending on the alignment of the rotation vector and magnetic field \citep{tsukamoto2015,wurster2016} –- although this scenario may be complicated by initial angular momentum distribution, magnetohydrodynamic structure, and turbulence (see review by \citealt{li2014}). 
Disk sizes for the other 20 disks in the full sample that do not show dust substructures are generally more compact.  A detailed comparison of disk sizes between the two sub-samples will be presented in our survey overview paper.

\subsection{Properties of Gaps and Rings}
In this section, we explore the general properties of gaps and rings revealed from our observations and their implications for disk properties and evolution. We analyze a total of 19 gap and ring pairs (e.g., a gap and the associated ring emission exterior to the gap) from our 12 disks. The very faint, outermost component of the MWC 480, GO Tau, and DL Tau disks are excluded in the analysis, since they were added to the fit to characterize the extended tenuous disk outer edge and do not necessarily represent a physical ring.  Four ring components that are not well resolved (the gap interior to the ring peak is not present), as seen in the model intensity profiles (see also Table~\ref{tab:gap_info}), are excluded from this analysis.  The inner cavities are also not considered in some of this analysis, since the gap locations cannot be well determined, although the gaps exterior to the inner ring are included.

\begin{figure}[!t]
\centering
    \includegraphics[width=0.45\textwidth]{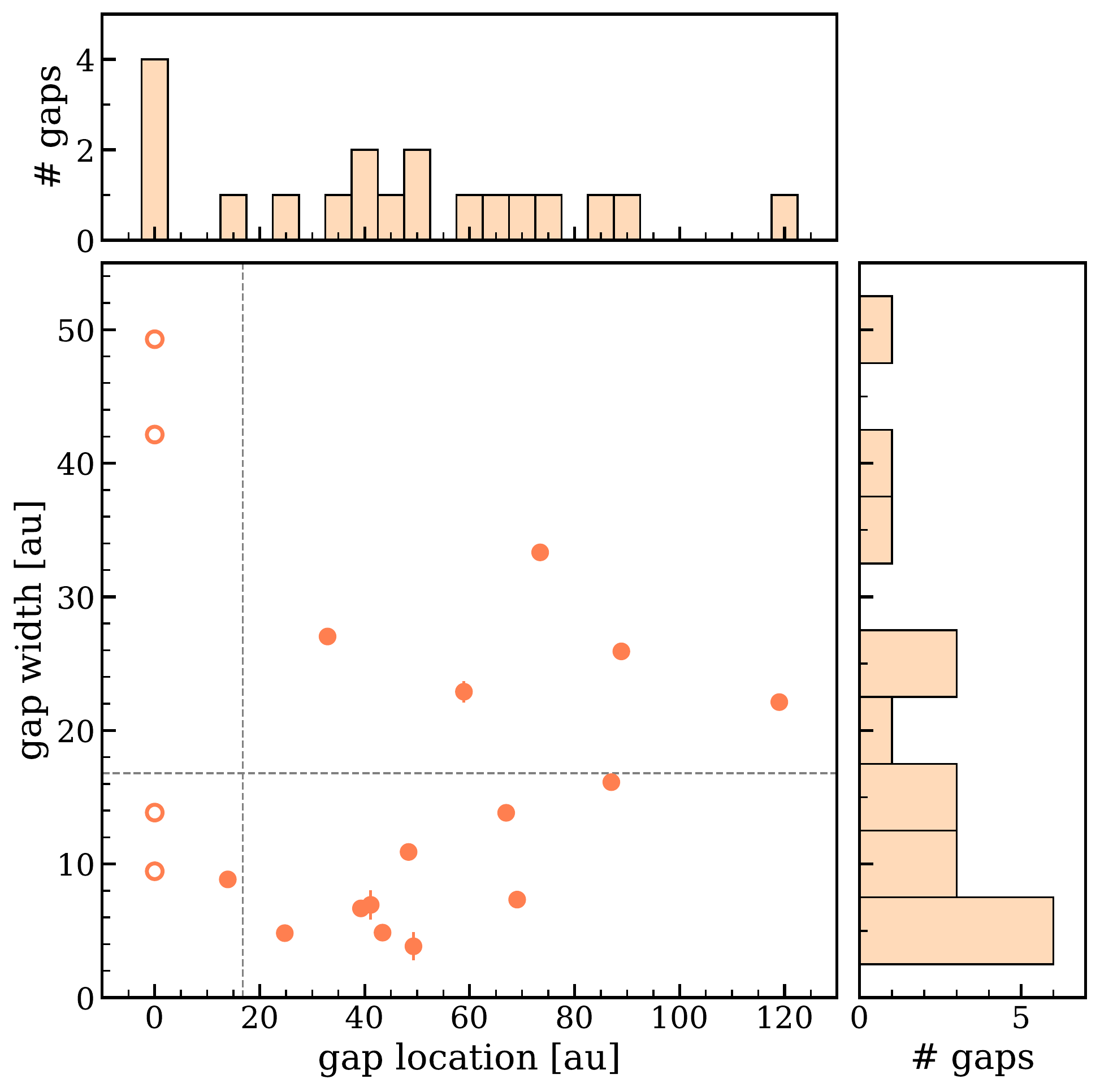}
    \caption{Gap location versus gap width for the 19 well-resolved gap and ring pairs (including the 4 inner cavities, with location set to 0 and shown as open circles; see the text for more details of which ring components are excluded). The grey dashed lines represent the typical beam size ($0.12''$), adopting a typical distance of 140 pc.  \label{fig:stat_gaps_v1}}
\end{figure}

Each substructure is described by the gap location, the gap width, and the intensity contrast ratio, as measured in our model fits.  The gap location is defined here as the radius where the intensity profile reaches a local minimum interior to the ring. The gap width is defined as the full width at half depth, in which the depth is the difference between the intensity at the gap location and the peak value of its outer ring.
We also measure the ring-gap contrast ratio as the intensity ratio at ring peak and gap location. The uncertainty for each parameter is estimated from the 16th-84th percentile (1$\sigma$) values from the chains of the last 1000 steps of our MCMC calculations.

\begin{figure}[!t]
\centering
    \includegraphics[width=0.4\textwidth]{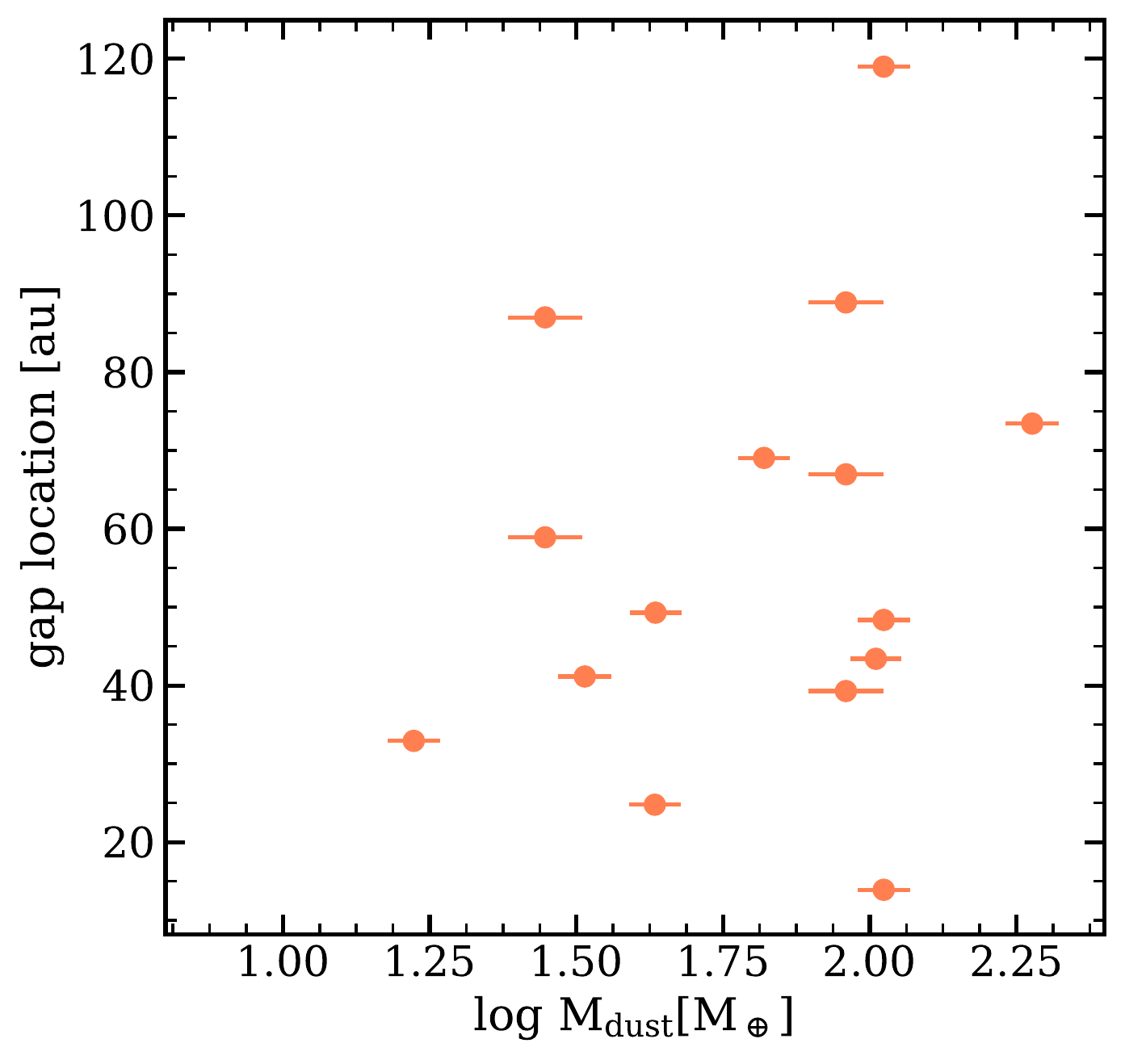}
    \caption{The gap location as a function of disk dust mass for 15 gaps (excluding the 4 inner cavities).  \label{fig:stat_4in1}}
\end{figure}

As shown in Figure~\ref{fig:stat_gaps_v1}, gaps are located from 10 to 120 au with no preferred distance. Most gaps are narrow and unresolved\footnote{If the gap width would have been measured as half of the distance between two adjacent rings, then a few very shallow gaps would have larger widths, but most gaps would still be unresolved or marginally resolved. Our conclusions related to gap widths would not be affected.}. With higher resolution, more substructures, narrower substructures, and lower contrast substructures would be expected to emerge (e.g., for TW Hydra, more rings were revealed in observations with $0\farcs02$ resolution observations \citep{andrews2016} than with $0\farcs3$ resolution \citep{zhang2016}). Narrow gaps around 100 au are absent in the current observations. A weak trend might be seen between gap location and gap width, in which gaps located further out have larger width, broadly consistent with the case of planet-disk interaction as will be discussed later in \S~\ref{sect:planet}. 
Moreover, as shown in Figure~\ref{fig:stat_4in1}, gap location does not depend on disk mass. We might see a desert of gaps located outside 40\,au for less massive disks, which seems to be consistent with the $L_{\rm mm} - R_{\rm eff}$ relation that fainter disks tend to be smaller in sizes \citep{tazzari2017,tripathi2017}. Alternatively, in order to retain a massive disk, substructures should be formed at larger radii, or the outer disk will be drained through fast inward drift. 

\begin{figure}[!t]
\centering
    \includegraphics[width=0.45\textwidth]{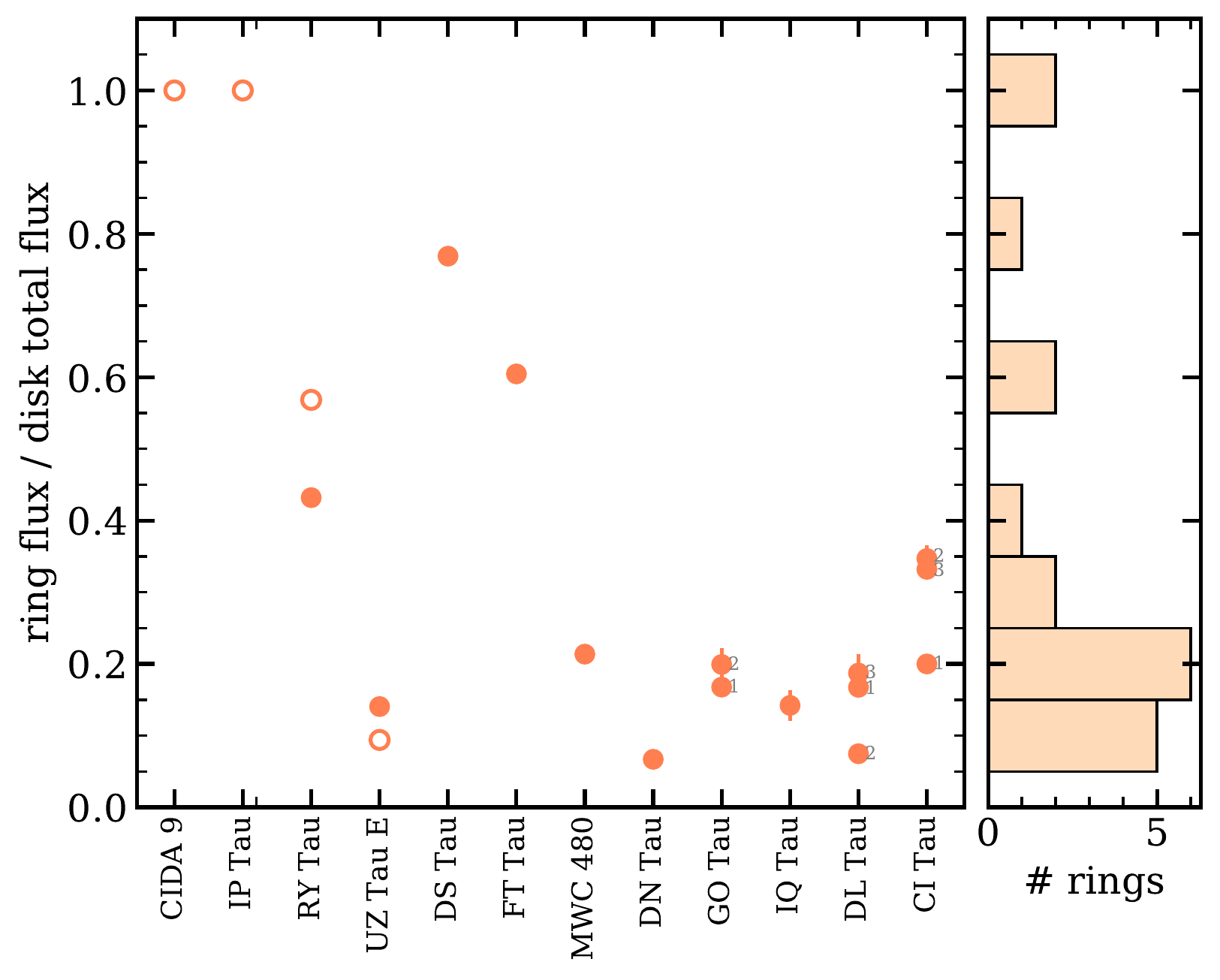}
    \caption{Left: The fractional flux of the disk in each ring. The four rings associated to inner cavities are shown as open circles. For disks with multiple rings, the numbers aside indicate the relative position of ring component from the star; Right: The histogram of fractional flux in the ring for 19 rings. 
    \label{fig:stat_gaps_v2}}
\end{figure}

Most ring-gap pairs have intensity contrast ratios lower than 3, with a few very depleted exceptions (the ring of DS Tau and MWC 480, the first ring of GO Tau) with ratios exceeding 20. Of the four inner cavities, IP Tau and CIDA 9 have nearly empty inner hole, while RY Tau and UZ Tau E only have a factor of two depletion. 
Figure~\ref{fig:stat_gaps_v2} illustrates the relative flux in each ring with respect to the total disk flux, which peaks around 0.2, with a tail towards higher fraction. Except for the two rings around inner cavities for IP Tau and CIDA 9,  the two rings in DS Tau and FT Tau (two single-ring disks) have more than 60\% of the total disk flux. The rings in disks with multiple substructures (e.g., CI Tau, DL Tau, GO Tau) generally hold $\sim$ 20\% of the total disk flux. This quantity is approximately proportional to the fraction of dust mass within the ring, though the optical depth of the dust and the temperature differences between the ring and the rest of the disk lead to substantial uncertainties. Since the back-reaction of dust on gas is strongest when dust-to-gas density ratio is of the order of unity \citep{youdin2005}, a typical 20\% accumulation of dust in the rings suggests that the creation of an individual ring in general may not be so relevant for the global disk dynamical evolution.  However, the total effect of all rings, including any rings in the inner disk that we could not detect and rings emerging from single-ring systems, may affect the dynamical evolution of the disk.

\subsection{Possible Origins for Gaps and Rings}
The exciting discovery of gap and ring-like features in the young HL Tau system \citep{HLTau2015ApJ} suggests that dust particles get trapped in local gas pressure bumps.  This and subsequent observations, including those presented here, have revealed that rings are prevalent in protoplanetary disks, with an importance that has motivated the development of many theoretical explanations of the observed substructures.  
Pressure bumps could be created outside the orbit of a planet (e.g., \citealt{pinilla2012}) or the outer edge of a low ionization region \citep[the so-called dead zone;][]{flock2015}. Other magneto-hydrodynamic effects, including zonal flows (e.g., \citealt{johansen2009}), could also play a role in gas evolution and gas pressure distribution.  Our focus of this section is to compare the rings and gaps in our sample to two popular hypothesis, that rings are carved by embedded planets or induced by condensation fronts, followed by future perspectives in discerning different mechanisms at play.

\subsubsection{Planet-disk Interactions} \label{sect:planet}
One of the widely invoked explanations for the observed gaps and rings in protoplanetary disks is related to the presence of embedded planets orbiting around the central star. The mass of the planet, the viscosity and pressure forces of the disk combine to determine the dynamical evolution of disk-planet interactions and thus the resulting distribution of the mm-sized dust grains. For the purposes of this subsection, we assume that the gaps are carved by planets and then use the ring and gap properties to estimate the masses of the hidden planets.  We then compare these results to statistics from exoplanet observations.

\begin{figure}[!t]
\centering
    \includegraphics[width=0.45\textwidth]{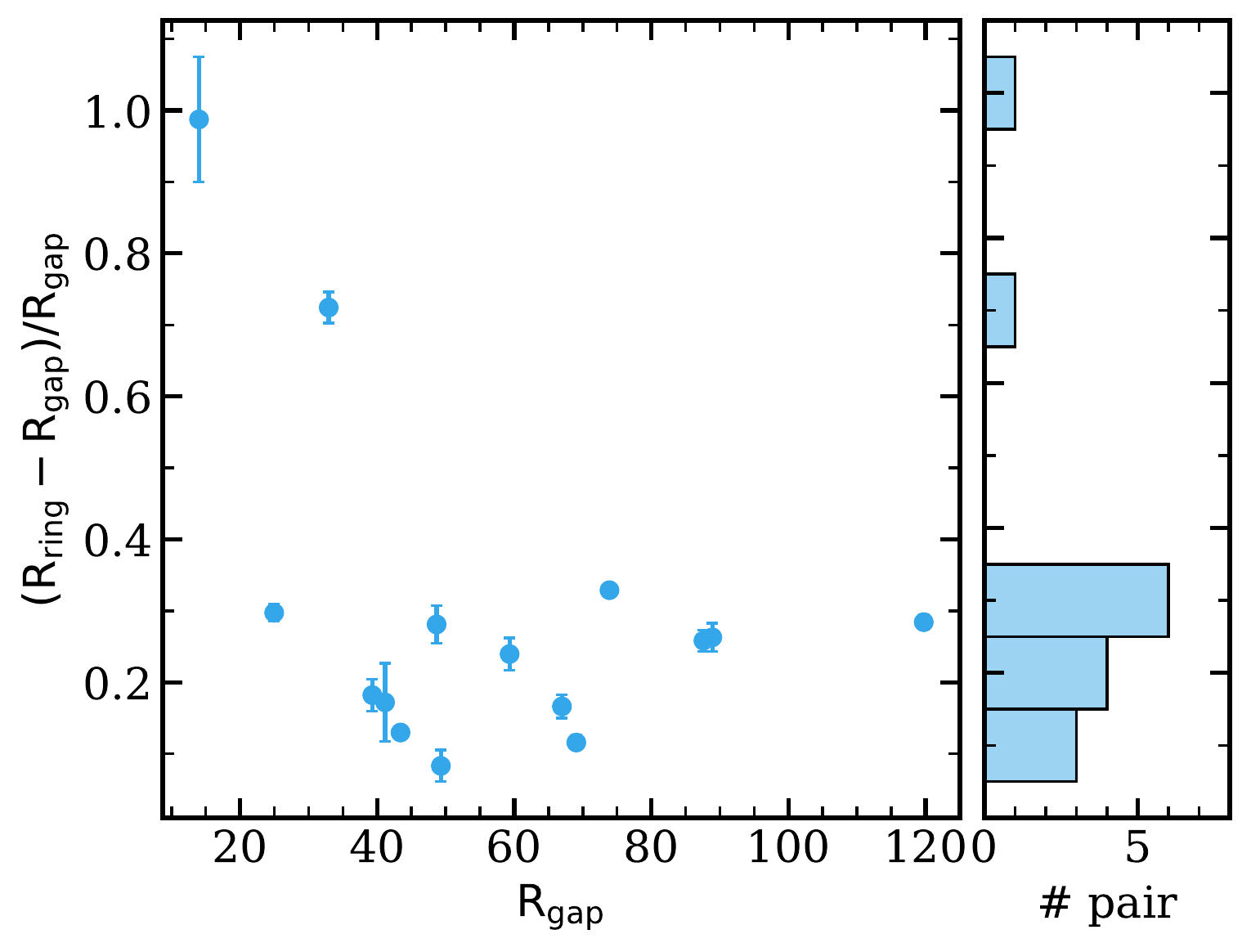} 
    \caption{Left: The gap-to-ring distance normalized to gap location (an indicator of planet mass) as a function of gap location; right: The histogram of the gap to ring distance.  
    \label{fig:stat_planet}}
\end{figure}

\begin{figure}[!t]
\centering
    \includegraphics[width=0.45\textwidth]{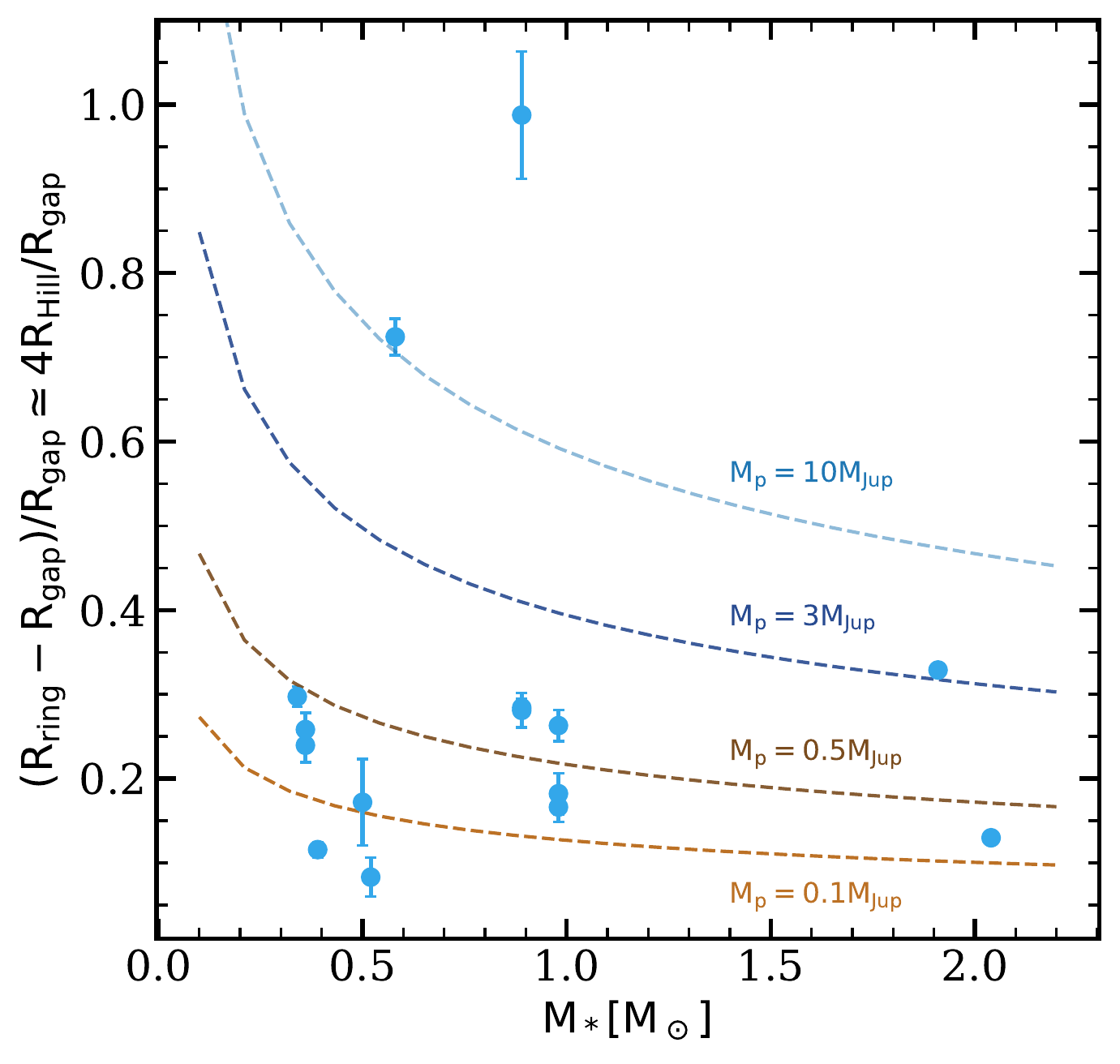} \\
    \caption{The gap-to-ring distance normalized to gap location (an indicator of planet mass) as a function of stellar mass. The dashed line represents how the planet mass indicator scales with stellar mass for a given planet mass, assuming gap-ring distance is 4 Hill radii. 
    \label{fig:stat_planet_Hill}}
\end{figure}

For planet-disk interactions, the gap location should occur at the orbital radius of the planet.  The ring of mm-size dust grows at the location of the local pressure maximum in the gas, outside the planet orbit.  A more massive planet will build a steeper pressure gradient, thereby forming a deeper and wider gap than would be created by a less massive planet \citep{fung2014,kanagawa2015,rosotti2016}.

The {\it minimum mass} of a planet that could form a gap in the gas density structure, leading to local pressure bump beyond the planet’s orbit, may be described analytically by
\begin{equation}
 \frac{M_{\rm p}}{M_{\star}}\propto \left(\frac{H}{r_{\mathrm p}}\right)^a \alpha^b,
 \label{eq:minimmass}
\end{equation}
where $r_{\rm p}$ is the distance of the planet to the star, $\alpha$ is the turbulence parameter, with power-law indices $a \in [2,3]$ and $b \in [0,1]$ \citep[see derivations in][]{lin1993,duffell2012,duffell2013,ataiee2018}.
If the disk is in vertically hydrostatic equilibrium ($H = c_{\rm{s}}/ \Omega_{\rm{k}}$), assuming uniform $\alpha$ and a power-law profile for the temperature ($T \propto r^{-1/2}$), the minimum planet mass able to create a pressure bump scales as $M_{\star}^{c} r_{\rm p}^{d}$ with $c \in [-1/2, 0]$ and $d\in [1/2,3/4]$. Assuming that the gap-ring distance (the distance of gap minimum and the ring peak, an alternative measurement for the gap width) scales with the Hill radius of the planet ($R_{\rm Hill} = r_{\rm p}(M_{\rm p}/(3M_{*}))^{1/3}$), and taking the planet masses given by Eq.~\ref{eq:minimmass}, the gap width normalized to the gap location is expected to scale as $ M_{\star}^{e} r_{\mathrm p}^{f}$ with $e \in [-1/3,-1/2]$ and $f\in [1/6,1/4]$, i.e., a weak dependence on both parameters. Figure \ref{fig:stat_planet} shows the distance between the ring peak and gap center, normalized to the gap location (presumably the location of any potential planet).  The normalized gap-ring distance is typically 0.2--0.3, with only two gaps as high diagnostic outliers (DS Tau and the closest gap of CI Tau). Given the lack of a clear trend between planet mass indicator and planet location, the mass of most planets in our sample (except for the outliers of CI Tau and DS Tau) might be close to the minimum planet mass able to produce a pressure bump beyond the planet orbit.

If we simply assume that the gap radius corresponds to 4\,$R_{\rm Hill}$ \citep{dodson-Robinson2011}, most of our gaps are related to planets with mass of 0.1--0.5\,$M_{\rm J}$, as shown in Figure~\ref{fig:stat_planet_Hill}. These estimated planet masses \footnote{In the contemporaneous study of CI Tau \citep{clarke2018}, hydrodynamic models of the gaps led to planets with masses of 0.15 and 0.4\,$M_{\rm J}$ for the outer two gaps, consistent with our simple estimation here.  The innermost planet is estimated here to be much more massive than 0.75\,$M_{\rm J}$ adopted in \citet{clarke2018}, with a difference likely driven by their ability to better resolve this gap.} 
have large uncertainties and should be interpreted as upper limits, since gap radius could extend to 7--10\,$R_{\rm Hill}$ (e.g., \citealt{pinilla2012}). An alternative way to estimate the mass of a planet associated with the gap is by linking the diagnostic of the gap-ring distance to hydrodynamic simulations \citep{rosotti2016}. The planet mass derived from this diagnostic highly depends on disk viscosity. When the turbulence parameter \citep{shakura1973} is assumed to be $\alpha=10^{-4}$, a value consistent with recent turbulence constraints by \citet{flaherty2015} and \citet{flaherty2018}, the diagnostic of 0.25 corresponds to a planet mass $\sim$ 15 M$_\oplus$ (0.05 M$_{\rm J}$). For a viscous accretion disk ($\alpha=10^{-2}$, \citealt{hartmann1998}), the related planet mass is about 0.3 M$_{{\rm J}}$ (\citealt{rosotti2016}; Facchini et al.~2018 in prep). The closest gap of CI Tau has the largest normalized gap-ring distance, corresponding to a Jupiter-mass planet in all cases. The large, spatially-resolved gap in DS Tau provides us with a hint that a massive planet may also reside in this disk.  These simulations were performed for a central star with 1 $M_\odot$ (\citealt{rosotti2016}; Facchini et al.~2018 in prep), thus the inferred planet mass should be re-scaled to the same planet/stellar mass ratio for stars with different masses. The estimated planet masses are subject to large uncertainties in disk properties defined in the models, including disk temperature, viscosity, and scale-height. This picture is also complicated by the ability of super-Earths to open multiple gaps in inviscid disks \citep{bae2017,dong2017,dong2018}. Dynamical interactions between a gaseous disk and an embedded planet could also produce spiral arms, which are not detected in our sample, probably because of the weak coupling of mm grains with the gas.

The analysis above excludes the four inner cavities. The infrared SEDs of these four systems indicate that small dust grains are still present in the inner disks, despite the depletion of mm-sized grain that is observed. Low mass planet(s) with traffic jam effects could be responsible for the cavity opening \citep{rosotti2016}.

The planets inferred from these analyses are challenging to compare to the statistics of known exoplanets.  Very few stars have massive planets at large radii, and even fewer planets are found around M dwarfs \citep[see review by][]{bowler2016}, however current sensitivities prevent the detection of the lower-mass planets that would create the gaps located at 20--100 au identified here.  Systems with super-Earths/sub-Jovian mass planets at radii larger than 30 au would be unlike our own solar system (ice giants of $\sim20M_\oplus$ at 20-30 au), but may be prevalent. The most common types of planets found in \textit{Kepler} transit and microlensing are super-Earths and Neptunes \citep{winn2015,suzuki2016,pascucci2018}, althrough the distribution at larger radii is not well constrained \citep{clanton2016,meyer2018}.

\begin{figure*}[!t]
\centering
    \includegraphics[width=0.99\textwidth]{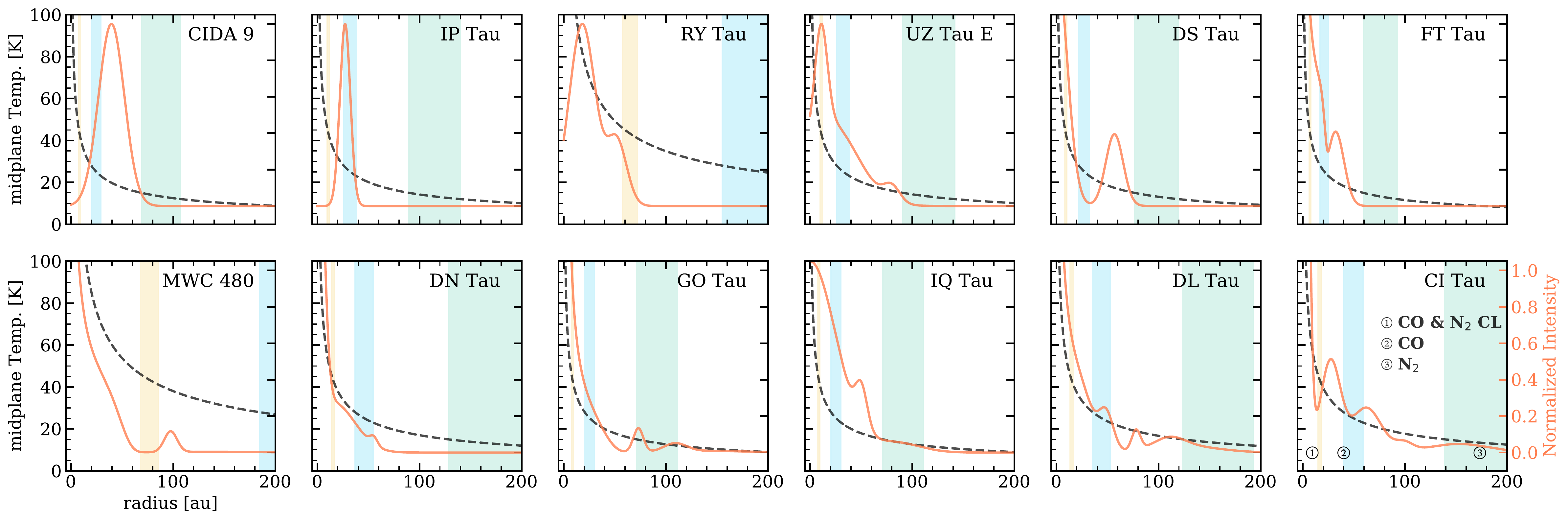}
    \caption{The disk midplane temperature profile (grey dashed line) and normalized intensity profile (orange line) for each disk. The matched ice line locations are marked as shaded regions, with N$_2$ in cyan, CO in blue, and the clathrate-hydrate CO and N$_2$ in orange. \label{fig:iceline_v2}}
\end{figure*}

\begin{figure}[!t]
\centering
    \includegraphics[width=0.45\textwidth]{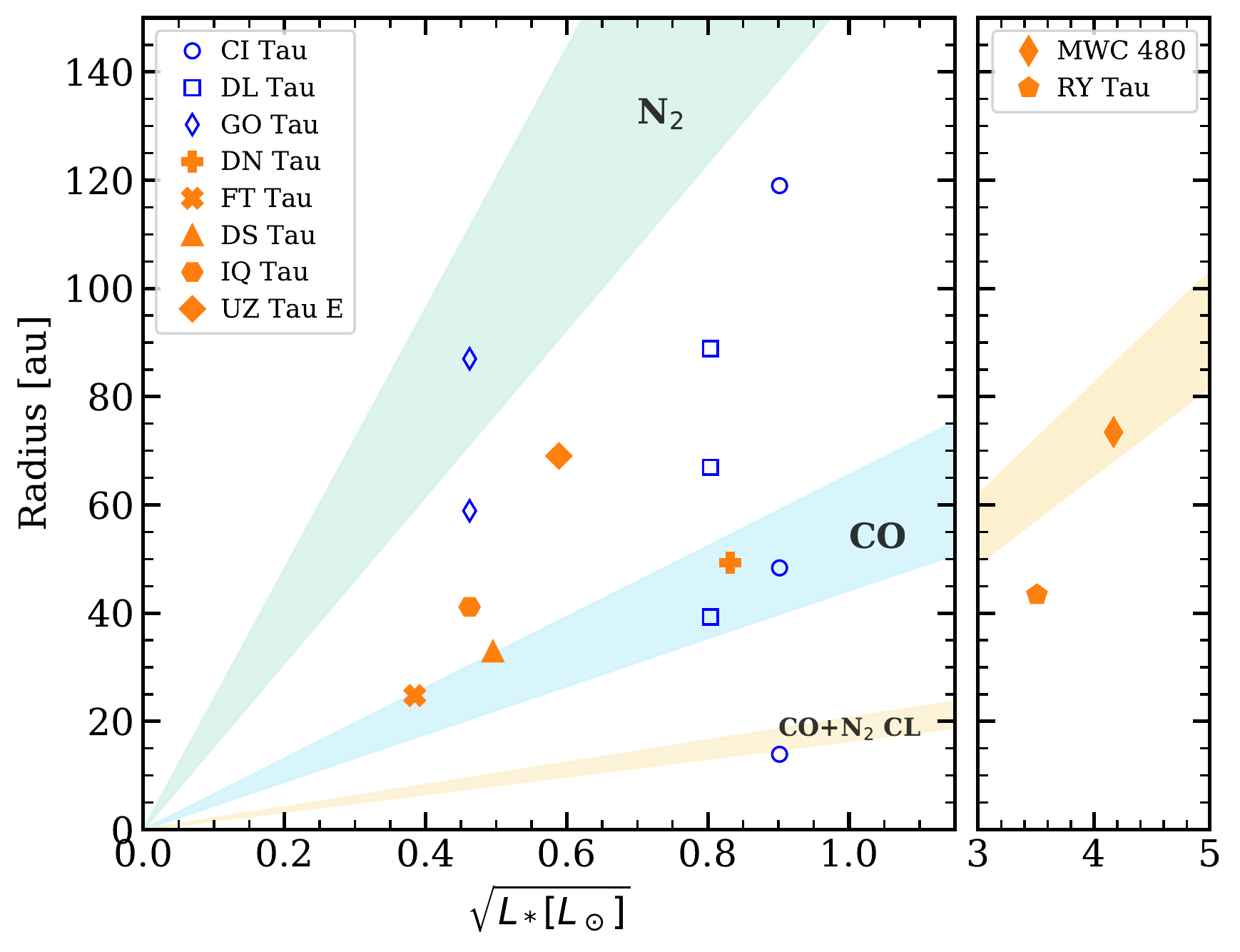}
    \caption{Disk radius versus the square root of stellar luminosity. The color-shaded regions correspond to the range of ice line location for N$_2$ (cyan), CO (blue), and the clathrate-hydrate CO and N$_2$ (orange), with condensation temperatures described in text and disk dust temperature approximated by Eq.~\ref{eq:temperature}. Gap locations for each disk are plotted in different symbols, and five gaps are located near CO ice line. 
    \label{fig:iceline_v1}}
\end{figure}

\subsubsection{Condensation Fronts} 
As disk temperature decreases towards larger radii, a series of major volatile molecules freeze out onto dust grains. These phase transition regions, also referred to as condensation fronts or ice lines, are locations where dust opacities and collisional growth/fragmentation are expected to change, producing features like gaps and rings seen in dust images (e.g., HL Tau, \citealt{zhang2015}, \citealt{banzatti2015}, \citealt{okuzumi2016}). If correct, this explanation should be universal and apply to all disks.

For an irradiated flared disk, \citet{kenyon1987} parameterize the disk midplane temperature as 
\begin{eqnarray}
T(r)&=&T_{\star}\left(\frac{R_{\star}}{r}\right)^{1/2} \phi_{\mathrm{inc}}^{1/4} ,
 \label{eq:temperature}
\end{eqnarray} 
where the flaring angle $\phi_{\mathrm{inc}}$ is assumed to be 0.05 \citep{dullemond2004}. 
The temperature profile for each disk is shown in Figure~\ref{fig:iceline_v2}. Taking the stellar luminosity
\footnote{In this paper, we use only the photospheric luminosity rather than the total luminosity, including the energy released by accretion.  The accretion luminosity is usually 1--10\% of the photospheric luminosity and can therefore be ignored, though in rare cases the accretion luminosity may dominate \citep[e.g.][]{manara17}. Accretion luminosity contributes as additional heating source mainly in the inner disk region, thus has small effects on the cold region we discussed in this paper.}  ($L_{\star}=4{\pi}R_{\star}^2{\sigma}{T_{\star}}^4$) into account, the ice line location for a specific species scales as $r \propto {L_{\star}}^{1/2}{T_{\rm{conden}}}^{-2} $. If gaps are formed around ice lines, the gap locations should occur in regions defined by major volatiles. Since the ice line regions associated with the abundant H$_2$O and NH$_3$ ice are unresolved for most of our observations (Figure~\ref{fig:iceline_v2}), our focus here is on molecules with lower condensation temperature: N$_2$ (12-15 K), CO (23-28 K), and clathrate hydrated CO and N$_2$ (41-46 K). The condensation temperatures, adopted from \citet{zhang2015}, correspond to the gas number densities of $10^{10}-10^{13} \rm{cm}^{-3}$ in disk midplane.

Figure~\ref{fig:iceline_v2} compares the location of gaps to the expected location of ice lines for each target. We also summarize the comparison in Figure~\ref{fig:iceline_v1} in the $r-\sqrt{L_*}$ plane. Five gaps lie close to the CO ice line (i.e. blue-shaded region matched with gap location), the closer-in gap of CI Tau and the gap of MWC 480 are located around clathrate hydrated CO+N$_2$ ice line, and the outer gap of GO Tau is located at a region with temperature consistent with N$_2$ ice line. The other seven gaps are unrelated to any of the ice lines we consider here. The four inner cavities are not included in this discussion, since temperature in the inner cavity could match with condensation temperatures of a series of major volatiles. For the disks with multiple, well-resolved gaps (CI Tau, DL Tau, and GO Tau), 1--2 gaps in each disk correspond to an ice line location, but 1--2 rings in each also do not match an ice line.  Among the gaps that are consistent with condensation fronts, explaining the wide and deep gaps of DS Tau and MWC 480 using condensation fronts alone would be challenging (see simulated images around ice lines in \citealt{pinilla2017}).

These comparisons between gap locations and snow lines depend on the disk temperature profile \citep{pinte2014} and the condensation temperatures for different species, both of which suffer from significant uncertainties.  
A detailed radiative transfer modeling on MWC 480 by Liu et al.~(2018) yields a midplane temperature for mm-sized grains that is a factor of 1.5 - 1.9 cooler than the parameterized temperature from Eq.~\ref{eq:temperature}.
A simple experiment of radiative transfer modeling\footnote{In order to derive the midplane temperature profile, we use the RADMC-3D code \citep{dullemond2012} to run radiative transfer models with steller properties fixed to the values given in Table~\ref{tab:source_prop} for each object. The details about the model setup can be found in Liu et al.~(2018). In general, the disk is assumed to be passively heated by the stellar irradiation, with two dust grain populations and mm flux fixed to our measurements. The flaring index and the scale height were set to 1.1 and 15 au, respectively, both of which are typical values found in multi-wavelength modeling of protoplanetary disks.} on the other 11 disks yields temperature profiles that are similar to analytic solutions (Eq.~\ref{eq:temperature}) beyond 20 au, where our gaps are located, for all sources except for the brighter objects, $\rm {RY\,Tau}$. Therefore, the comparison of iceline locations to gap locations is not significantly affected.

The temperature range for volatile condensation also varies under different conditions and depends on the dust grain size, composition, and surface area that the molecule freezes onto. Taking CO ice as an example, \citet{oberg2011} suggest an average condensation temperature of 20 K for CO ice, lower than the value adopted from \citet{zhang2015}, due to a different assumption for the gas number density in the disk midplane. The disk region corresponding to the CO ice line would then move outward. The specific gaps that are associated with CO ice lines would change, but their total number would not increase.

The behavior of dust particles around ice lines is currently not well understood. From a physical point of view, inward drifting particles lose their surface ice when crossing the condensation fronts, causing a higher dust-to-gas ratio just outside the ice line, and the evaporated gas may diffuse outward and re-condense onto dust outside the ice line, both leading to enhanced grain growth beyond the ice line \citep{cuzzi2004,stammler2017}. Differential grain growth inside and outside the ice lines, producing spectral index variations, has been suggested around H$_2$O ice line \citep{banzatti2015}, with observational evidence from the outburst system V883 Ori \citep{cieza2016}.  Any change in spectral index of dust emission has not yet been detected for the CO ice line \citep{stammler2017}. Suppressed grain growth for dust particles outside the ice lines due to sintering could also cause pile-ups of dust in the region where smaller grains have lower drift velocity \citep{okuzumi2016}. In addition, \citet{zhang2015} suggested rapid grain growth at radii corresponding to different ice condensation fronts, reducing the detectable mm flux in the gaps. The nature of ice sublimation and condensation near these ice lines is complex, because of the strong dependence on the evolution of dust aggregates in terms of fragmentation/coagulation for dust with different compositions, as well as radial drift and turbulent mixing overall \citep{pinilla2017}.

Given the physical and chemical complexity of the problem, reflected in the model uncertainties described above, and the absence of a clear correlation and correspondence between the gap locations and ice lines, it is still very challenging to test and reconcile the observed substructures with the locations of different ice lines. Volatile condensation fronts of CO, N$_2$ and clathrate hydrated CO+N$_2$ may not be a universal solution for all observed gaps and rings for our sample, but could play a role in shaping some disk dust structures.

\subsubsection{Distinguishing Different Mechanisms of Gap Creation}

Distinguishing between the different mechanisms responsible for gap and ring formation will yield a better understanding of planet formation, either by providing an indirect probe of planets or by providing a diagnostic of physics that would likely be important in the growth of planetesimals. 
The rings and gaps in our sample have a wide range of properties. The radial location of the rings do not seem to prefer the expected location of snow lines, and the gaps of DS Tau and MWC 480 are likely too wide to be explained by ice lines. Around ice lines, the composition of particles are altered and hence their aerodynamical behavior with the gas \citep{pinilla2017}, but without significant changes of the gas density profile, and therefore pressure bumps are not created.

On the other hand, the variety of ring properties might be expected for hidden exoplanets, which would occur at a wide range of locations and masses.  Unfortunately, the planet mass function and their radial distributions are not yet well enough known at such low masses to be able to test whether Neptune-mass planets are prevalent at these distances.

Comparing dust morphology for different grain sizes observed at different wavelength will allow us to distinguish particle trapping from the other mechanisms (e.g., ice lines). Additional tests of these and other theories will emerge with better maps, including the distribution of gas in the disk. These observations require deep integrations but are feasible for small numbers of disks as follow-ups from surveys such as ours.  Indeed, CO cavities and gaps have been reported in a few bright disks (e.g., \citealt{isella2016}; \citealt{fedele2017}; \citealt{boehler2018}), lending to some evidence for planet-disk interaction.  However, some degeneracy is introduced because the CO gaps could be interpreted as either a gap in the gas density or in the thermal structure \citep{facchini2017}.

\section{Conclusions} 		\label{sect:conclusions}
This paper presents the analysis of dust substructures detected at 1.33 mm continuum emission from 12 disks, observed at $\sim 0.12''$ resolution in an ALMA Cycle 4 survey of 32 disks in Taurus star forming region. Rings and gaps are the most common type of substructure in our selected disks and exhibit a wide variety of properties. Disk model fitting is performed in the visibility plane to quantify the amplitude, location and width for each substructure component. We then study the stellar and disk properties for the selected sample in the context of dust morphology, and the origins of dust substructures from analysis of the gap and ring properties. Our main findings are summarized as follows.

\begin{enumerate}
\item  The 12 disks with detected substructures span a wide range in stellar mass and disk brightness. Disks with multiple rings tend to be more massive and more extended (larger effective radius) than those with single rings.

\item Four disks are identified with inner dust cavities with a radius of 5--25 au and different levels of depletion. The IR SEDs reveal the presence of small dust grains at where large grains are depleted, consistent with expectations for dust filtration by a low mass planet. Three of these four disks are relatively massive, consistent with expectations from the $M_* - M_{\rm dust}$ relationship for disks with cavities.  These disks may be a collection of heterogeneous sources; more observations towards the fainter end might help to understand the origins of disk inner cavities.

\item We resolve 19 gap-ring pairs in the model intensity profiles of the 12 disks.  The locations, sizes, and contrasts of these gap-ring pairs are used to investigate how these features form. Dust gaps are located from 10 to 120 au.  Most gaps have narrow widths that are smaller than the beam size ($\sim16$ au). A typical fraction of 20-30\% of mm fluxes are accumulated in each ring, with a few exceptions.

\item The presence of wider gaps at larger radii hints for planet-disk interaction. The low intensity contrast in most ring and gap pairs suggests the possible link to low mass planets. We follow the diagnostic used in planet-disk interaction simulations (the separation of ring and gap normalized to gap location) to infer planet mass, and find that super-Earths and Neptunes are good candidates if disk turbulence is low ($\alpha=10^{-4}$), in line with the most common type of planets discovered so far.

\item We do not observe a concentration of gap radii around major ice line locations. While five gaps are located close to the expected radii of CO ice lines, and another one and two gaps are related to N$_2$ and clathrate hydrated CO+N$_2$ ice lines, several other gaps have no relationship with the estimated radii of any ice line we considered.  Forming gaps and rings around condensation fronts may not be a universal explanation for all our observed substructures, but could play some role in shaping some of the disks. If ice lines cause rings, it remains unclear why condensation fronts would affect only some of the disks.
\end{enumerate}

Multi-wavelength observations that probe different sizes of dust grains will help to discern particle trapping (e.g., planets) from other mechanisms (e.g., ice lines), with the expected narrower dust accumulation at longer wavelength if dust particles are trapped in rings. Characterizing the gas distribution across the gaps will also be essential to better constrain their origins. Follow-up observations in these forms are timely and will accelerate our understanding of disk evolution and planet formation.

\paragraph{Acknowledgments}
\acknowledgments{We thank the anonymous referee for useful comments.  FL thanks Sean Andrews for discussions on comparing ALMA fluxes to past interferometer fluxes, and Zhaohuan Zhu, Kaitlin Kratter and Andrew Youdin for insightful discussions.  FL and GJH thank Subo Dong for discussions of planet mass functions and radial distributions.    
FL and GJH are supported by general grants 11473005 and 11773002 awarded by the National Science Foundation of China.  PP acknowledges support by NASA through Hubble Fellowship grant HST-HF2-51380.001-A awarded by the Space Telescope Science Institute, which is operated by the Association of Universities for Research in Astronomy, Inc., for NASA, under contract NAS 5-26555. GD acknowledges financial support from the European Research Council (ERC) under the European Union's Horizon 2020 research and innovation programme (grant agreement No 681601). DH is supported by European Union A-ERC grant 291141 CHEMPLAN, NWO and by a KNAW professor prize awarded to E. van Dishoeck. CFM acknowledges support through the ESO Fellowship. MT has been supported by the DISCSIM project, grant agreement 341137 funded by the European Research Council under ERC-2013-ADG. FM, GvdP and YB acknowledge funding from ANR of France under contract number ANR-16-CE31-0013 (Planet-Forming-Disks). DJ is supported by NRC Canada and by an NSERC Discovery Grant. BN and GL thank the support by the project PRIN-INAF 2016 The Cradle of Life - GENESIS-SKA (General Conditions in Early Planetary Systems for the rise of life with SKA). YL acknowledges supports by the Natural Science Foundation of Jiangsu Province of China (Grant No. BK20181513) and by the Natural Science Foundation of China (Grant No. 11503087). 
This paper makes use of the following ALMA data: 2016.1.01164.S. ALMA is a partnership of ESO (representing its member states), NSF (USA) and NINS (Japan), together with NRC (Canada), MOST and ASIAA (Taiwan), and KASI (Republic of Korea), in cooperation with the Republic of Chile. The Joint ALMA Observatory is operated by ESO, AUI/NRAO and NAOJ. This work has made use of data from the European Space Agency (ESA) mission Gaia (https://www.cosmos.esa.int/gaia), processed by the Gaia Data Processing and Analysis Con- sortium (DPAC, https://www.cosmos.esa.int/web/gaia/dpac/consortium). Funding for the DPAC has been provided by national institutions, in particular the institutions participating in the Gaia Multilateral Agreement.
} 

\software{Galario \citep{tazzari2018}, emcee \citep{Foreman-Mackey2013}, RADMC-3D \citep{dullemond2012}, CASA (v5.1.1; \citealt{McMullin2007})}




\clearpage
\appendix

\section{Fitting results for individual disk} \label{sec:note}

\setcounter{figure}{0}
\renewcommand{\thefigure}{A\arabic{figure}}

\begin{figure}[ht]
\centerline{\includegraphics[scale=0.9,trim=0 50 0 50]{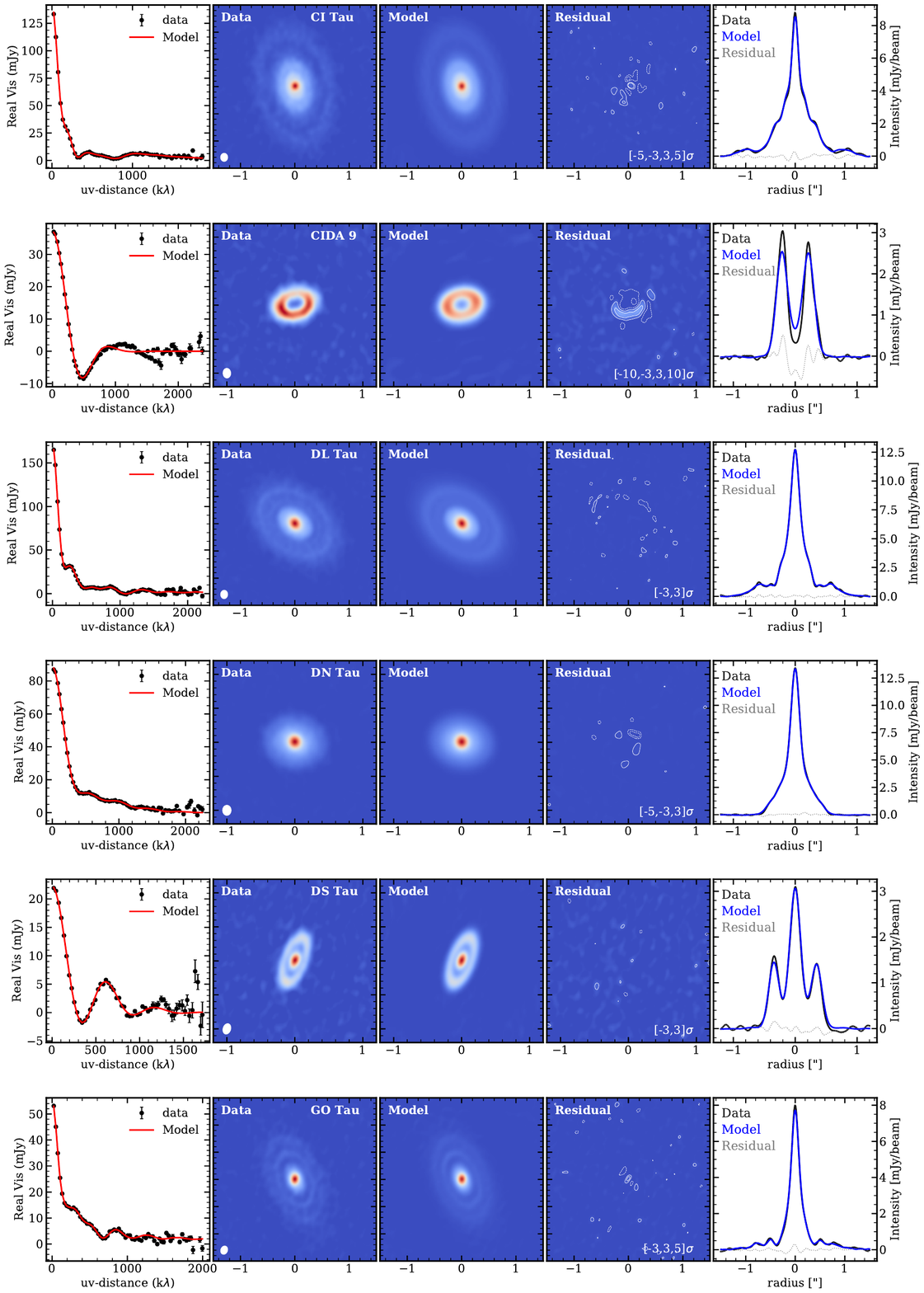}}
\captcont{A comparison of data and best-fit model for individual disk, including binned and deprojected visibility profile, continuum images (data, model, and residual maps), and radial profile along the disk major axis. \label{fig:model_result_all}}
\end{figure}

\begin{figure}[ht]
\centerline{\includegraphics[scale=0.9,trim=0 50 0 50]{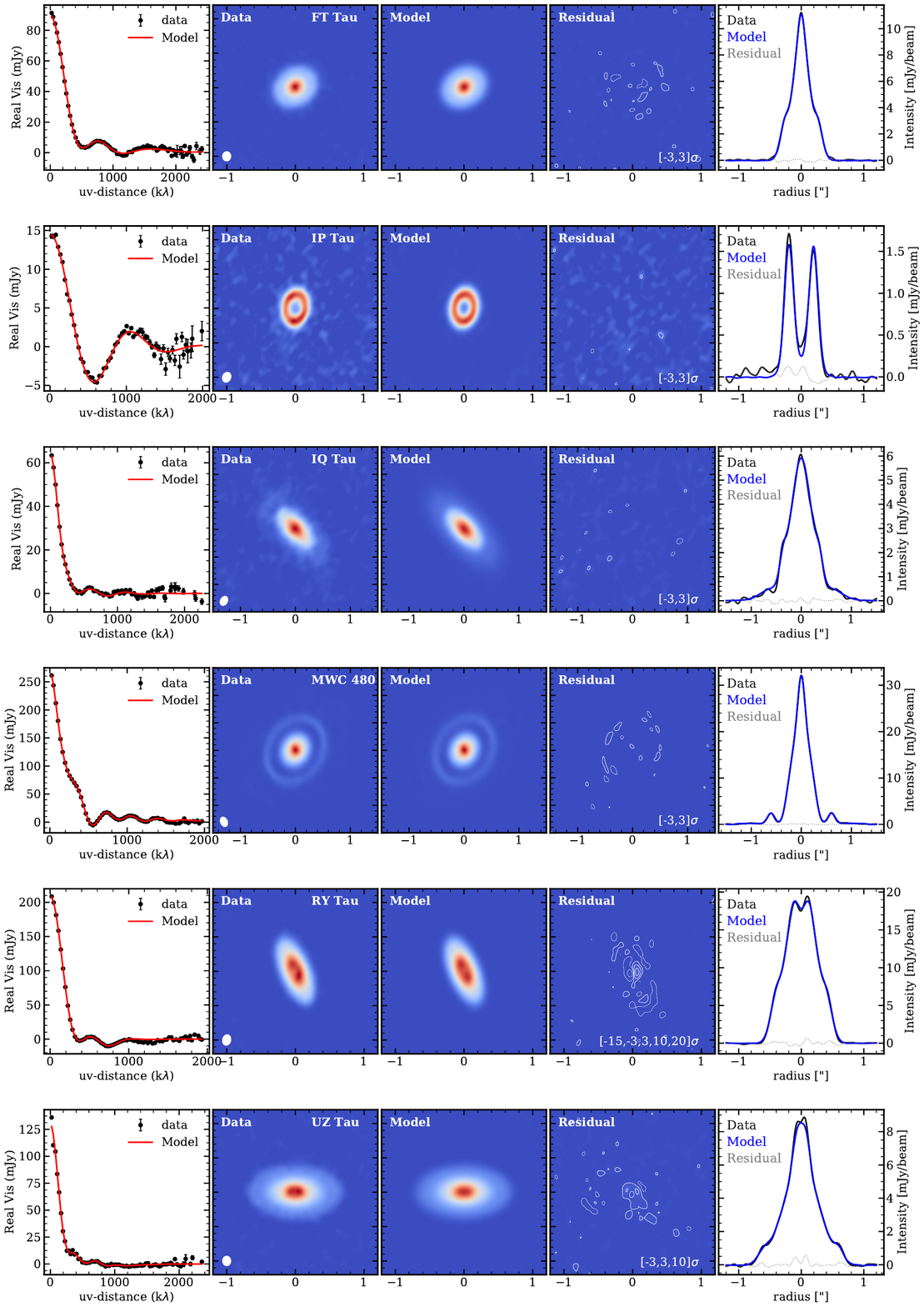}}
\captcont{Cont.}
\end{figure}


\begin{deluxetable*}{lcccccc}[!t]
\tabletypesize{\scriptsize}
\tablecaption{Source Properties and observation results\label{tab:gap_info}}
\tablewidth{0pt}
\tablehead{
\colhead{Name} & \colhead{Ring Number} & \colhead{Ring Location} & \colhead{Ring Width} & \colhead{Gap Location} & \colhead{Gap Width}  & \colhead{contrast} \\
\colhead{} & \colhead{} & \colhead{(au)} &\colhead{(au)} & \colhead{(au)} &  \colhead{(au)} & \colhead{} 
} 
\startdata
CIDA 9    &  1 & 39.52  $\pm$  0.00   & 25.31  $\pm$ 0.00   &                  --  & 49.28  $\pm$ 0.17 &                -- \\  
IP Tau    &  1 & 27.05  $\pm$  0.07   & 10.40  $\pm$ 0.13   &                  --  & 42.15  $\pm$ 0.13 &                -- \\  
RY Tau    &  1 & 18.19  $\pm$  0.00   & 25.60  $\pm$ 0.13   &                  --  & 13.85  $\pm$ 0.13 &                -- \\  
          &  2 & 49.04  $\pm$  0.14   & 19.46  $\pm$ 0.13   &  43.41   $\pm$ 0.13  &  4.86  $\pm$ 0.20 &  1.03 $\pm$  0.01 \\  
UZ Tau E    &  1 & 11.02  $\pm$  0.13   & 12.05  $\pm$ 0.26   &                  --  &  9.46  $\pm$ 0.13 &                -- \\  
          &  2 & 17.42  $\pm$  1.31   & 28.16  $\pm$ 0.66   &                  --  &               --  &                -- \\ 
          &  3 & 77.04  $\pm$  0.47   & 15.72  $\pm$ 0.26   &  69.05   $\pm$ 0.20  &  7.34  $\pm$ 0.46 &  1.07 $\pm$  0.02 \\  
DS Tau    &  1 & 56.78  $\pm$  0.16   & 17.17  $\pm$ 0.16   &  32.93   $\pm$ 0.32  & 27.03  $\pm$ 0.24 & 24.07 $\pm$  1.87 \\  
FT Tau    &  1 & 32.14  $\pm$  0.13   & 16.51  $\pm$ 0.13   &  24.78   $\pm$ 0.19  &  4.83  $\pm$ 0.06 &  1.37 $\pm$  0.01 \\  
MWC 480   &  1 & 97.58  $\pm$  0.08   & 12.56  $\pm$ 0.16   &  73.43   $\pm$ 0.16  & 33.33  $\pm$ 0.16 & 73.78 $\pm$ 13.00 \\  
DN Tau    &  1 & 15.36  $\pm$  0.77   & 21.12  $\pm$ 0.38   &                  --  &               --  &               --  \\ 
          &  2 & 53.39  $\pm$  0.95   &  7.68  $\pm$ 0.64   &  49.29   $\pm$ 0.44  &  3.84  $\pm$ 1.21 &  1.06 $\pm$  0.09 \\  
GO Tau    &  1 & 73.02  $\pm$  0.16   &  9.79  $\pm$ 0.43   &  58.91   $\pm$ 0.66  & 22.90  $\pm$ 0.86 & 17.83 $\pm$  5.21 \\  
          &  2 & 109.45 $\pm$  0.36   & 22.18  $\pm$ 0.86   &  86.99   $\pm$ 0.88  & 16.13  $\pm$ 0.58 &  4.54 $\pm$  0.63 \\  
IQ Tau    &  1 & 48.22  $\pm$  1.09   & 11.79  $\pm$ 0.92   &  41.15   $\pm$ 0.63  &  6.94  $\pm$ 1.29 &  1.10 $\pm$  0.08 \\  
          &  2 & 82.79  $\pm$  2.88   & 24.50  $\pm$ 1.96   &                  --  &               --  &               --  \\ 
DL Tau    &  1 & 46.44  $\pm$  0.48   & 14.63  $\pm$ 0.48   &  39.29   $\pm$ 0.32  &  6.68  $\pm$ 0.48 &  1.09 $\pm$  0.03 \\  
          &  2 & 78.08  $\pm$  0.24   &  8.59  $\pm$ 0.64   &  66.95   $\pm$ 0.87  & 13.83  $\pm$ 0.72 &  6.36 $\pm$  1.32 \\  
          &  3 & 112.27 $\pm$  0.32   & 29.57  $\pm$ 1.27   &  88.90   $\pm$ 1.11  & 25.92  $\pm$ 0.56 &  2.11 $\pm$  0.12 \\  
CI Tau    &  1 &  27.67 $\pm$  0.24   & 19.28  $\pm$ 0.47   &  13.92   $\pm$ 0.32  &  8.85  $\pm$ 0.16 &  2.20 $\pm$  0.12 \\  
          &  2 &  61.95 $\pm$  0.47   & 29.39  $\pm$ 1.11   &  48.36   $\pm$ 0.41  & 10.90  $\pm$ 0.40 &  1.22 $\pm$  0.04 \\  
          &  3 &  99.22 $\pm$  1.58   &  8.69  $\pm$ 0.95   &                  --  &                -- &                -- \\  
          &  4 & 152.80 $\pm$  0.47   & 59.72  $\pm$ 0.47   &  118.99  $\pm$ 0.65  & 22.12  $\pm$ 0.55 &  1.67 $\pm$  0.05 \\  
\enddata
\tablecomments{Ring and gap properties for each disk. Gap properties and intensity contrasts are blank for the gaps that are not resolved in model intensity profiles.}
\end{deluxetable*}

\end{CJK*}
\end{document}